\documentclass{article}
\usepackage{graphicx}
\usepackage[all]{xy}

\pdfoutput=1

\usepackage{jheppub}
\usepackage{amsmath}
\usepackage{xfrac}
\usepackage{bm}
\usepackage{epsfig,multicol,bbm}
\usepackage{url}

\usepackage{graphicx}
\usepackage{setspace}
\usepackage{amssymb}
\usepackage{listings}
\usepackage{epstopdf}
\usepackage{float}
\usepackage{color}
\usepackage{amsmath}
\usepackage{ dsfont }
\usepackage{slashed}
\usepackage{cancel}
\usepackage{subfigure}

\newcommand{\newc}{\newcommand}
\newc{\gsim}{\lower.7ex\hbox{$\;\stackrel{\textstyle>}{\sim}\;$}}
\newc{\lsim}{\lower.7ex\hbox{$\;\stackrel{\textstyle<}{\sim}\;$}}
\newc{\gev}{\,{\rm GeV}}
\newc{\mev}{\,{\rm MeV}}
\newc{\ev}{\,{\rm eV}}
\newc{\kev}{\,{\rm keV}}
\newc{\tev}{\,{\rm TeV}}

\newc{\mz}{M_Z}
\newc{\mpl}{M_*}
\newc{\mw}{m_{\rm weak}}
\newc{\nr}[1]{N^c_R{}_{#1}}
\usepackage{amsmath}

\newcommand{\tr}{\textrm{tr}}

\newcommand{\btheta}{\bar{\theta}}
\newcommand{\qDelta}{\widetilde{\Delta}}
\newcommand{\qf}{\widetilde{f}}
\newcommand{\qM}{\widetilde{M}}

\newcommand*\xbar[1]{%
  \hbox{%
    \vbox{%
      \hrule height 0.5pt 
      \kern0.2ex
      \hbox{%
        \kern-0.1em
        \ensuremath{#1}%
        \kern-0.05em
      }%
    }%
  }%
} 

\def\beq{\begin{equation}}
\def\eeq{\end{equation}}
\def\bea{\begin{eqnarray}}
\def\eea{\end{eqnarray}}
\def\bitem{\begin{itemize}}
\def\eitem{\end{itemize}}
\newc{\ie}{{\it i.e.}}          \newc{\etal}{{\it et al.}}
\newc{\eg}{{\it e.g.}}          \newc{\etc}{{\it etc.}}
\newc{\cf}{{\it c.f.}}

\renewcommand{\binom}[2]{\left(\begin{matrix}
#1 \\
#2
\end{matrix}   \right)}

\newcommand\fverb{\setbox\fverbbox=\hbox\bgroup\verb}
\newcommand\fverbdo{\egroup\medskip\noindent%
            \fbox{\unhbox\fverbbox}\ }
\newcommand\fverbit{\egroup\item[\fbox{\unhbox\fverbbox}]}
\newbox\fverbbox

\newcommand{\Nm}[1]{{\!N\!-\!#1}}

\newcommand\T{\rule{0pt}{2.6ex}}       
\newcommand\B{\rule[-1.2ex]{0pt}{0pt}} 

\title{Scherk-Schwarz Supersymmetry Breaking in 4D}

\date{\today}

\author[a]{Nathaniel Craig,}
\author[b]{Hou Keong Lou,}

\affiliation[a]{Department of Physics and Astronomy \\
 Rutgers University \\
Piscataway, NJ 08854 }
\affiliation[b]{Department of Physics \\ Princeton University \\ Princeton, NJ 08540}

\preprint{RUNHETC-2014-10}    

\abstract{ Using the techniques of dimensional deconstruction, we present 4D models that fully reproduce the physics of 5D supersymmetric theories compactified on an $S^1/\mathbb{Z}_2$ orbifold with general Scherk-Schwarz supersymmetry breaking (SSSB) boundary conditions. In contrast to previous approaches, our deconstruction involves only soft supersymmetry breaking. Deconstruction preserves many of the attractive features of SSSB without the cumbersome architecture of a full fifth dimension, ambiguity of the ultraviolet completion, or challenges associated with stabilizing a large radius of compactification. We proceed to deconstruct various five-dimensional models featuring Scherk-Schwarz boundary conditions, including folded supersymmetry.

}


\begin{document}

\maketitle

\section{Introduction}

The discovery of an apparently elementary Standard Model-like Higgs at the LHC \cite{ATLAS-CONF-2014-009, CMS-PAS-HIG-13-005} has brought the hierarchy problem sharply to the fore, while the absence of evidence for conventional signs of naturalness at the LHC renders the solution of the hierarchy problem increasingly unclear. This tension between naturalness and LHC limits is highly generic, whether the hierarchy problem is solved by supersymmetry or compositeness. All viable solutions to the hierarchy problem with a high (i.e, $\gtrsim$ TeV) cutoff require additional symmetries to protect the Higgs mass from ultraviolet sensitivity. The $\mathcal{O}(1)$ top yukawa coupling requires the top quark to transform under these symmetries as well and, since these symmetries typically commute with the Standard Model gauge group, this implies the existence of top partners charged under QCD. A suite of LHC searches for such top partners has pushed limits out to $\sim 700$ GeV after the first run of the LHC \cite{ATLAS-CONF-2013-065,ATLAS-CONF-2013-037, ATLAS-CONF-2013-024, Chatrchyan:2013iqa, Chatrchyan:2013xna,Chatrchyan:2013uxa}, imperiling the radiative stability of the weak scale. This tension is exacerbated in supersymmetric theories with a high cutoff, as the stop mass is radiatively connected to the gluino mass and direct gluino limits are yet stronger by almost a factor of two \cite{Aad:2014pda, Aad:2013wta, Chatrchyan:2013wxa, Chatrchyan:2014lfa}.

There are various opportunities to relax the tension between direct search limits and naturalness of the weak scale (for a recent review, see \cite{Craig:2013cxa}). One simple option in the context of supersymmetry is for the stops to decay into an LSP with a small relative mass splitting, typically erasing the decisive missing energy signature \cite{LeCompte:2011cn}. Another option is to radically lower the cutoff in supersymmetric scenarios or endow the gluino with a Dirac mass so that the gluino and stop can be parametrically separated \cite{Kribs:2012gx}. A third, and more radical option, is to arrange for the natural top partners to be neutral under QCD, as can happen when symmetries protecting the Higgs mass are broken by orbifold projection. The orbifolded daughter theory does not possess the full symmetries of the parent theory, but scalars in the daughter theory are nonetheless protected by the parent symmetry at one loop \cite{Kachru:1998ys, Lawrence:1998ja, Bershadsky:1998mb, Kakushadze:1998tr, Bershadsky:1998cb, Schmaltz:1998bg}. Realizations of this loophole include the twin Higgs model \cite{Chacko:2005pe} and folded supersymmetry \cite{Burdman:2006tz}, both of which feature top partners neutral under QCD thanks to an orbifolded symmetry and preserve naturalness in the face of LHC limits. 

All three of these options for reconciling naturalness with LHC limits are naturally realized in five-dimensional theories on an interval where symmetries are reduced by boundary conditions. In particular, gauge-, global-, and super-symmetries can all be reduced by Scherk-Schwarz symmetry breaking boundary conditions \cite{Scherk:1978ta}, and a modest radius of compactification can provide a low cutoff to the four-dimensional effective theory. Such theories have long been fruitfully exploited for novel phenomenology \cite{Pomarol:1998sd, Antoniadis:1998sd, Delgado:1998qr, Barbieri:2000vh, Barbieri:2001dm} and have recently attracted renewed interest for features such as typically compressed spectra \cite{Murayama:2012jh}, relative insensitivity to radiative corrections \cite{Dimopoulos:2014aua}, genericness of Dirac gauginos, and freedom to decouple fermionic superpartners of the Higgs \cite{Dimopoulos:2014aua}. Moreover, the Scherk-Schwarz mechanism provides an avenue to realize exotic theories with a natural weak scale, such as an explicit construction of folded supersymmetry \cite{Burdman:2006tz}. 

However, all five-dimensional theories are inevitably accompanied by a certain degree of baggage, including full Kaluza-Klein (KK) towers of four-dimensional matter. Especially for theories with supersymmetry, multiplets of KK towers need to be added in order to respect the $\mathcal{N}_{5D}=1$ supersymmetry in five dimensions. Another problem with five-dimensional theories is that they are not ultraviolet (UV) complete, and they require a cutoff that is often not much larger than the inverse compactification radius $1/R$. In order to stabilize the weak scale, a low cutoff will demand quantum gravity to enter at tens of TeV, resulting in very little parametric freedom consistent with bulk symmetries. In addition, stabilizing the compactification radius in Scherk-Schwarz theories with a small cosmological constant requires intricate engineering \cite{Ponton:2001hq}, as well as an additional layer of dynamics to accommodate potential recent evidence for inflationary tensor perturbations \cite{Ade:2014xna}.

While theories with Scherk-Schwarz supersymmetry breaking (SSSB) offer promising new venues for naturalness, 5D complications strongly motivate reproducing the same physics in 4D via dimensional deconstruction \cite{ArkaniHamed:2001ca} of a supersymmetric extra dimension \cite{ArkaniHamed:2001ed, Csaki:2001em, Csaki:2001qm, Cheng:2001an, Kobayashi:2001fr, Falkowski:2002gx, Iqbal:2002ep, Cohen:2003xe, Dudas:2003iq, DiNapoli:2006kc}. Scherk-Schwarz breaking of gauge and global symmetries have previously been successfully deconstructed \cite{Csaki:2001em, Csaki:2001qm}. Several proposals exist for the reproduction of SSSB in four dimensions \cite{ArkaniHamed:2001ed, Falkowski:2002gx}. However, as we will see, these proposals fail to capture the full physics of SSSB boundary conditions and suffer from UV sensitivities. In this paper we present the first complete four-dimensional deconstruction of five-dimensional theories on an $S^1/\mathbb{Z}_2$ orbifold subject to general Scherk-Schwarz twist of an $SU(2)_R$ symmetry. We depart from previous approaches by exploiting  the five-dimensional relationship between Scherk-Schwarz boundary conditions and radion-mediated supersymmetry breaking, explicitly deconstructing the latter rather than the former.\footnote{Although there has been a previous attempt to deconstruct radion mediation \cite{Kobayashi:2001fr}, our results and conclusions differ considerably.} Thus our deconstruction involves only soft breaking of supersymmetry, protecting the theory against UV sensitivity.

Our paper is organized as follows: in Section \ref{sec:ss} we begin by reviewing the spectrum of five-dimensional supersymmetric theories compactified on a circle or an $S^1/\mathbb{Z}_2$ orbifold, with Scherk-Schwarz twists of an $SU(2)_R$ symmetry. We then discuss previous attempts at deconstructing Scherk-Schwarz physics and discuss their shortcomings. Afterward, we review the relationship between SSSB and radion mediation. In Section \ref{sec:decon} we review various aspects of supersymmetric deconstruction, including the agreement between 4D and 5D physics at the level of both the action and the mass spectrum. We then turn to the deconstruction of radion mediation in Section \ref{sec:radion}, explicitly reconstructing the action and mass spectrum of 5D orbifold theories with SSSB. We explicitly deconstruct several canonical 5D examples in Section \ref{sec:ex} before concluding in Section \ref{sec:conc}. We reserve several technical details for the appendices, including the explicit deconstruction of the 5D action for a non-abelian gauge theory in Appendix \ref{app:a}, the complete diagonalization of mass matrices for the Scherk-Schwarz deconstruction in Appendix \ref{app:b}, and a review of the boundary conditions for folded supersymmetry in Appendix \ref{app:c}.

\section{The Path to Scherk-Schwarz in 4D} \label{sec:ss}

We begin by reviewing the spectrum arising from compactification of a flat fifth dimension on a $S^1/\mathbb{Z}_2$ orbifold with the Scherk-Schwarz twist of an $SU(2)_R$ symmetry.\footnote{For a systematic discussion, see \cite{Barbieri:2001dm}.} While there are several existing proposals \cite{ArkaniHamed:2001ed, Falkowski:2002gx} for four-dimensional theories with analogous spectra, these proposals rely on hard breaking of supersymmetry. Although the ensuing scalar mass spectrum depends on this hard breaking only at high loop order, such theories nonetheless suffer from UV sensitivities associated with the absence of supersymmetry at high scales. Moreover, as we will see, these proposals only reproduce a discrete subset of possible Scherk-Schwarz spectra.

\subsection{Scherk-Schwarz in 5D}

Consider a $\mathcal{N}_{5D} =1$ supersymmetric gauge theory in five dimensions with gauge group $G$ and at least one bulk hypermultiplet. From the four-dimensional perspective the theory possesses two $\mathcal{N}=1$ supersymmetries.\footnote{We adopt the convention that $\mathcal{N}$ refers to amount of supersymmetry in four dimension, and `multiplet' refers to 4D vector or chiral multiple unless explicitly specified.} The 5D vector multiplet contains the on-shell components $\bm{V} = (A_\mu, \lambda_1, \lambda_2, \Sigma)$, while the 5D hypermultiplet contains on-shell components $\bm{\Phi} = (\Phi, \Phi^c, \psi, \psi^c)$. The bulk Lagrangian exhibits an $SU(2)_R$ symmetry under which the supercharges of the two supersymmetries, the gauge fermions $\lambda_1, \lambda_2$, and the matter scalars $\Phi, \Phi^{c\dagger}$ each form doublets. 

We can compactify the fifth dimension with radius $R$ by identifying $x_5 + 2 \pi R$ with $x_5$. The dimensional reduction of bulk 5D fields gives rise to a tower of KK states, where the $n^\textrm{th}$ states all have mass $m_n = n/R$ and form $\mathcal{N} = 2$ multiplets. In particular, the $\Sigma^{(n)}$ scalar is eaten to render the vector $A_\mu^{(n)}$ massive, while the gauge fermions $\lambda_1^{(n)}, \lambda_2^{(n)}$ and the matter fermions $\psi^{(n)}, \psi^{c(n)}$ respectively combine to form Dirac fermions.  

This $\mathcal{N} = 2$ tower leads to a phenomenologically unviable set of vector-like zero modes. One solution is to compactify the extra dimension on an  $S^1/\mathbb{Z}_2$ orbifold. This orbifold is equivalent to carrying out the above circular compactification and imposing a discrete $\mathbb{Z}_2$ parity identifying $x_5$ with $-x_5$, acting on fields $\Psi$ as $\Psi(x^\mu, x^5) \to \pm \Psi(x^\mu, -x^5)$. By assigning different parity to different fields in the same 5D multiplet, the orbifold breaks one of the two $\mathcal{N} = 1$ supersymmetries and removes half of the original tower of KK modes. The $n \neq 0$ modes arrange themselves into $\mathcal{N}=2$ multiplet as before. The zero modes must have even parity, and thus preserve only an $\mathcal{N} = 1$ supersymmetry. By assigning an even parity only to $(A_\mu, \lambda_1)$ and $(\Psi,\psi)$, the zero mode contains only an $\mathcal{N}=1$ vector and chiral multiplet. 

Finally, we can break all of the 4D supersymmetries by imposing on the 5D fields a nontrivial global-symmetry rotation under $2 \pi R$ translations, acting on the fields as $\Psi(x^\mu, x_5 + 2 \pi R) = e^{2 \pi i \alpha T} \Psi(x^\mu, x^5)$; this is the Scherk-Schwarz twist, which breaks the symmetry generated by $T$. For supersymmetry breaking, one candidate generator is $T = \sigma_2$ for the $SU(2)_R$ symmetry. This choice splits the $n$th Dirac gaugino into two Majorana fermions of mass $|n \pm \alpha|/R$, and the $n$th KK hypermultiplet complex scalars into two complex scalars of mass $|n \pm \alpha|/R$. The mass spectrum of the $n$th KK gauge boson, adjoint scalar, and hypermultiplet fermions are untouched as they do not transform under $SU(2)_R$. For $|\alpha| <\sfrac{1}{2}$, the lightest gauginos and hypermultiplet scalars have masses $\alpha/R$, and supersymmetry breaking is controlled by two parameters $1/R$ and $\alpha/R$. 

The specific choice $\alpha = \sfrac{1}{2}$ is noteworthy. As $\alpha \to \sfrac{1}{2}$ the Majorana fermions from adjacent KK levels coincide, such that the zero mode spectrum consists only of a massless gauge boson and chiral fermion, while Majorana gauginos again pair up to form Dirac multiplets. In particular, the $n$th KK towers include Dirac gauginos and complex scalars of mass $|n - \sfrac{1}{2}|/R$ alongside the usual spectrum of  KK gauge bosons, adjoint scalars, and matter fermions.

\subsection{Twisted supersymmetry and its discontents}

Our objective is to reproduce the physics of Scherk-Schwarz supersymmetry breaking in a purely four-dimensional framework through dimensional deconstruction \cite{ArkaniHamed:2001ca}. A four-dimensional analogue of SSSB called ``twisted supersymmetry'' was proposed in \cite{ArkaniHamed:2001ed} by introducing hard supersymmetry breaking to an otherwise supersymmetric quiver. In \cite{ArkaniHamed:2001ed} the physics of a supersymmetric extra dimension are reproduced by a conventional supersymmetric quiver, albeit with explicit supersymmetry breaking in the form of relative phases between the gauge-link field and the gaugino-link field couplings. Such phases can always be removed locally at a given node in the quiver by field redefinitions, but in a circular quiver there is one physical phase -- the global phase -- that is invariant under field redefinitions. 
The global phase $\theta$ leads to a shift in the tree-level spectrum of gauge bosons and fermions, namely
\begin{align} \label{eq:twistedspectrum}
m_{B,n}^2 &= \left( \frac{2}{a} \right)^2 \sin^2 \left( \frac{n a}{2R} \right) \\ \nonumber
m_{F,n}^2 &=  \left( \frac{2}{a}\right)^2 \sin^2 \left( \frac{n a}{2R} + \theta \right)
\end{align}
where $a$ is the lattice spacing.

This appears to reproduce the physics of SSSB for modes lighter than $1/a$, but upon closer inspection it differs in two key respects. The first discrepancy is evident in the mass spectrum. For $\theta = 0, \sfrac{1}{2}$ one reproduces the spectra of a Scherk-Schwarz theory with $\alpha = 0, \sfrac{1}{2}$, but for any other value of $\theta$ there is no correspondence with a Scherk-Schwarz spectrum. In particular, while the phase $\theta$ can take arbitrary values as one expects for generic Scherk-Schwarz boundary conditions, the mass eigenstates are always Dirac -- in contrast with genuine Scherk-Schwarz theories, where the gauge fermions are Majorana for general $\alpha$. The twisted construction {\it cannot} be simply altered to fix the spectra, since introducing hard supersymmetry breaking into the gaugino-link field coupling will only ever lead to a spectrum of Dirac fermions.

The second discrepancy is that the phase $\theta$ is only physical if the quiver is circular, such that a global phase is well-defined. Breaking the circular quiver by removing a link field or gauge node allows the independent re-phasing of all couplings, restoring supersymmetry. However, an unbroken quiver gives rise to an $\mathcal{N}=2$ spectrum of zero modes, as is apparent in Equation \ref{eq:twistedspectrum}; the bosonic zero modes include both a massless gauge boson and a massless adjoint scalar. Therefore we encounter an immediate tension between retaining a physical phase and reproducing the physics of a realistic $S^1/\mathbb{Z}_2$ orbifold.

An alternate proposal for Scherk-Schwarz-like physics in four dimensions entails breaking supersymmetry explicitly at a node by removing states from supermultiplets by hand \cite{Falkowski:2002gx}. The Lagrangian is also augmented with a mass term external to the quiver dynamics in order to reproduce the Scherk-Schwarz spectrum. As in the case of twisted supersymmetry, this can only reproduce the spectrum of an $\alpha = \sfrac{1}{2}$ Scherk-Schwarz theory, as the spectrum of KK gauginos is always Dirac. The advantage of this procedure is compatibility with $S^1/\mathbb{Z}_2$ orbifold deconstruction since it does not rely on global phases. The disadvantage is that quadratic sensitivities to the cutoff typically arise at two loops since the couplings on the non-supersymmetric node are no longer related by supersymmetry.

In these proposals, locality in theory space protects scalars from quadratic sensitivity to the cutoff to two-loop order \cite{Falkowski:2002gx} or $N$-loop order \cite{ArkaniHamed:2001ed}. Beyond that, these theories are still intrinsically UV sensitive; supersymmetry is ultimately not a good symmetry of the global quiver. One must either legislate that explicit supersymmetry breaking is restricted only to a contrived set of possible terms -- although in general the potential giving rise to link field vevs is no longer protected -- or furnish a UV completion for the apparent hard breaking. This strongly motivates the construction of UV-complete four-dimensional theories capable of producing the full range of Scherk-Schwarz breaking spectra with arbitrary $\alpha$, without reliance on arbitrary explicit supersymmetry breaking.

\subsection{From Scherk-Schwarz to radion mediation}

In order to capture the physics of arbitrary Scherk-Schwarz supersymmetry breaking twists without hard breaking terms, we will exploit the relationship between Scherk-Schwarz and radion-mediated supersymmetry breaking. As shown in \cite{Marti:2001iw}, the two mechanisms are related by a field redefinition. The identification makes transparent the UV insensitivity of SSSB by relating it to a manifestly soft breaking of supersymmetry. To mimic the physics of SSSB in four dimensions, we will exploit this equivalence and explicitly deconstruct a 5D theory where supersymmetry is broken by an $F$-term expectation value for the radion.

To motivate the four-dimensional setup we first review the 5D equivalence between radion mediation and Scherk-Schwarz supersymmetry breaking, following \cite{Marti:2001iw, Kaplan:2001cg}.  In five dimensions, the off-shell $\mathcal{N}_{5D}=1$ vector supermultiplet can be written in terms of two $\mathcal{N}=1$ superfields: a vector superfield $V$ and an adjoint chiral superfield $\chi$. In Wess-Zumino gauge, we have
\begin{equation}
V= -\theta \sigma^\mu \btheta A_\mu - i\btheta^2\theta \lambda_1 + i \theta^2\btheta\bar{\lambda}_1 + \frac{1}{2}\btheta^2\theta^2 D \qquad \chi = \frac{1}{\sqrt{2}}(\Sigma+iA_5)+\sqrt{2}\theta\lambda_2 + \theta^2 F_\chi
\end{equation}
For simplicity, consider the case of a $U(1)$ gauge theory. The 5D action contains only two terms,
\begin{equation}
S=\frac{1}{g_5^2}\int d^5x\left\{\frac{1}{4}\int d^2\!\theta \,  W^\alpha W_\alpha +  \int d^4 \!\theta\;  \left[\partial_5 V - \frac{1}{\sqrt{2}}(\chi^\dagger_i + \chi_i) \right]^2\right\}
\end{equation}

Ultimately, we will be interested in the 5D theory compactified on a circle with radius $R$. It is therefore useful to re-parametrize the coordinate $x_5 = R\phi$ and rescale our coupling $g_5^2 \rightarrow g_4^2/2\pi R$, upon which the action becomes
\begin{equation}
S=\frac{1}{g_4^2}\int d^4x\int \frac{d\phi}{2\pi}\left\{\frac{1}{4}\int d^2\!\theta \,W^\alpha  W_\alpha +  \int d^4 \!\theta\; \frac{1}{R^2}\left[\partial_\phi V - \frac{R}{\sqrt{2}}(\chi^\dagger_i + \chi_i) \right]^2\right\} .
\label{eq:5D_abelian_action}
\end{equation}

The generalization to a non-abelian gauge group can be constructed by adding appropriate factors of $e^V$, namely
\begin{equation}
S=\frac{1}{g_4^2}\int d^4x\int \frac{d\phi}{2\pi}\left\{\frac{1}{2}\int d^2\!\theta \,\tr\,W^\alpha  W_\alpha + \int d^4 \!\theta\;  
\frac{1}{R^2} \tr\left[e^{V}\partial_\phi e^{-V} + \frac{R}{\sqrt{2}}(\chi^\dagger_i + e^{-V}\chi_ie^V) \right]^2\right\}
\label{eq:5D_nonabelian_action}
\end{equation}

We can also add matter fields to the theory in the form of 5D hypermultiplets, which we can write in terms of $\mathcal{N}=1$ chiral superfields $\Phi, \Phi^c$ transforming in the (anti)fundamental representation. The matter hypermultiplet has the following action in terms of the re-parameterized coordinates:
\begin{equation}
S_{\textrm{matter}}=\int d^4x\int \frac{d\phi}{2\pi}\left[
 \int d^4 \!\theta\; \, \Phi^\dagger e^V \Phi  +
\int d^4 \!\theta\; \, \Phi^{c} e^{-V} \Phi^{c\dagger} +
\int d^2 \!\theta\; \frac{1}{R} \, \Phi^c \left(\partial_\phi + \frac{R}{\sqrt{2}}\chi \right)\Phi + \textrm{h.c.}
\right] .
\label{eq:5D_matter_action}
\end{equation}

Thus far we have treated the compactification radius $R$ as a constant. To include coupling with the radion, we include the radion chiral superfield $T$, with components
\begin{equation}
T=\frac{1}{R}\left(R+ iB_5 + \theta \Psi^5 + \theta^2 F_T\right)
\end{equation}
where $B_5$ and $\Psi^5$ are the graviphoton and gravitino in the $x_5$ extra dimension. The action is modified by making the replacements
\begin{eqnarray}
\int d^4 \!\theta\; \frac{1}{R^2}\left[\partial_\phi V - \frac{R}{\sqrt{2}}(\chi^\dagger_i + \chi_i) \right]^2 &\longrightarrow&
\int d^4 \!\theta\; \frac{1}{R^2}\frac{2}{T+T^\dagger}\left[\partial_\phi V - \frac{R}{\sqrt{2}}(\chi^\dagger_i + \chi_i) \right]^2 \notag \\
\int d^4 \!\theta\; \Phi^\dagger e^V \Phi &\longrightarrow&
\int d^4 \!\theta\; \frac{T+T^\dagger}{2}\, \Phi^\dagger e^V \Phi \notag \\
\int d^4 \!\theta\; \Phi^c e^{-V} \Phi^{c\dagger} &\longrightarrow&
\int d^4 \!\theta\; \frac{T+T^\dagger}{2}\, \Phi^c e^{-V} \Phi^{c\dagger} \notag \\
\frac{1}{2g^2}\int d^2\!\theta \,\tr\, W^\alpha W_\alpha  &\longrightarrow&
\frac{1}{2g^2}\int d^2\!\theta \,T\,\tr\, W^\alpha W_\alpha 
\label{eq:radion_replacement}
\end{eqnarray}

Note that the original action is recovered with $T=1$. Without the superfield $T$, terms of the form $\int d^4\theta\, \chi^{\dagger2}$ vanish identically, and the Lagrangian can be rewritten in a variety of ways up to boundary terms. However, when all the components of the superfield $T$ are included, additional terms of the form $\int d^4\theta\, \chi^{\dagger2}/(T+T^\dagger)$ induce nonvanishing couplings with the radion and cannot be neglected. This is particularly important when supersymmetry breaking is communicated through a non-vanishing $F_T$ expectation value.

Let us now demonstrate the correspondence between radion mediation and Scherk-Schwarz boundary conditions. In the gauge sector, a non-vanishing $F_T$ induces Majorana masses for the fermions,
\begin{equation}
\mathcal{L}_{\cancel{\textrm{SUSY}}} \ni 
- \frac{F_T}{4R} \,\lambda^a_1\lambda^a_1 + 
\frac{F^\dagger_T}{4R} \,\lambda^a_2\lambda^a_2
\end{equation}
Together with the fermion kinetic terms, these soft masses can be written in a manifestly $SU(2)_R$ invariant way through the combination $\bm{\lambda}_i =(\lambda_1, -i\lambda_2)$:
\begin{equation}
\mathcal{L} \ni i\bm\lambda_i^{a\dagger} \bar\sigma^\mu \partial_\mu \bm\lambda_i^a + 
\frac{1}{2R}\bm\lambda_i^a \epsilon_{ik}\left(\delta_{kj}\partial_\phi +i f_{kj}\right) \bm\lambda_j^a 
\qquad
f_{ij}=
\frac{1}{2}
\left(
\begin{matrix}
0 & -iF_T^\dagger \\
iF_T & 0 
\end{matrix}
\right)
\label{eq:gaugino_mass_continuum}
\end{equation}

Written in this way, the soft supersymmetry breaking masses can be absorbed by the field redefinition $\bm\lambda_i \rightarrow e^{i\phi f_{ij}}\bm\lambda_j$, explicitly demonstrating the correspondence between a  Scherk-Schwarz phase and radion mediation. Analogous arguments go through for additional matter $\Phi,\Phi^c$; a non-zero $F_T$ mixes their scalar and auxiliary components,
\begin{eqnarray}
\mathcal{L}_{\cancel{\textrm{SUSY}}} \ni 
\frac{F_T}{2R}\, (F^\dagger_{\Phi}\Phi + F^\dagger_{\Phi^c}\Phi^c) + \textrm{h.c.} 
\end{eqnarray}
This mixing modifies the values of the auxiliary components $F_{\Phi},F_{\Phi^c}$ when they are integrated out, such that
\begin{eqnarray}
-F^\dagger_{\Phi^c} &\longrightarrow& \frac{1}{R} 
\left[ \left(\partial_{\phi} -\frac{i}{2}RA_5 \right) - \frac{1}{2}R\Sigma \right] \Phi + \frac{F_T}{2R} \Phi^{c\dagger}
\notag\\
F_{\Phi} &\longrightarrow& \frac{1}{R} 
\left[ \left(\partial_{\phi} -\frac{i}{2}RA_5 \right) + \frac{1}{2}R\Sigma \right] \Phi^{c\dagger} - \frac{F^\dagger_T}{2R} \Phi
\end{eqnarray}
To make the $SU(2)_R$ symmetry manifest, we can define $\bm{\Phi}_i=(\Phi, \Phi^{c\dagger})$ and $D_\phi=(\partial_\phi -iRA_5/2)$. The resulting kinetic terms for the $\Phi,\Phi^c$ fields are
\begin{eqnarray}
-\left|D_\mu \bm{\Phi}_i\right|^2 + \frac{1}{R^2}\left|(D_{\!\phi}\delta_{ij}+if_{ij})\, \bm{\Phi}_j\right|^2 + \frac{1}{4}\left|\Sigma \,\bm{\Phi}_i\right|^2
\qquad
f_{ij}=
\frac{1}{2}
\left(
\begin{matrix}
0 & -iF_T^\dagger \\
iF_T & 0 
\end{matrix}
\right)
\end{eqnarray}
As with the vector multiplet, we see that the soft  breaking can be absorbed into the redefinition $\bm{\Phi}_i \rightarrow e^{i\phi f_{ij}}\bm{\Phi}_i$. Consequently, the mass spectrum of radion mediation agrees with the mass spectrum of Scherk-Schwarz breaking with an $R$-symmetry twist upon the identification $F_T  = 2\alpha$.

The advantage for realizing the Scherk-Schwarz mechanism as radion mediation is that supersymmetry breaking by a radion $F$-term is manifestly soft and under good UV control. This immediately suggests that the deconstruction of SSSB can proceed entirely through soft supersymmetry-breaking terms by adding analogous radion couplings, avoiding the potential UV sensitivity of previous proposals and capturing all possible values of the Scherk-Schwarz twist $\alpha$. This is the route we will take in Section \ref{sec:radion}.

\section{The Supersymmetric Deconstruction} \label{sec:decon}
In this section we review key concepts for supersymmetric deconstruction, with special emphasis on the analogue between 4D and 5D physics, in terms of both the action and the geometric boundary conditions. As we will see, there are many subtleties involved in correctly reproducing the physics of an extra dimension. The original deconstruction for a non-supersymmetric theory is presented in \cite{ArkaniHamed:2001ca}; we will only be considering supersymmetric theories and our discussion will largely follow the discussion in \cite{Csaki:2001em}.

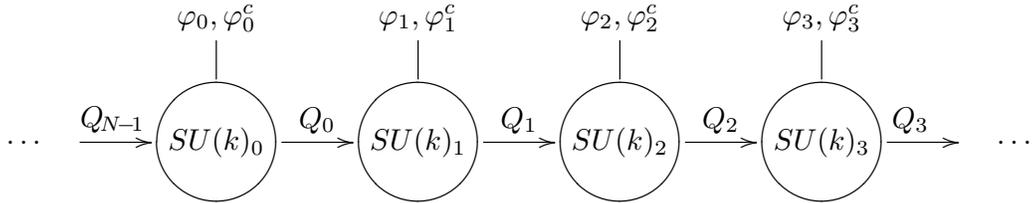
\begin{figure}[htb!]
\begin{equation*}
\resizebox{14cm}{!}{
\xymatrix{
&
{\varphi_0,\varphi_0^c}\ar@{-}[d(0.5)] &
{\varphi_1,\varphi_1^c}\ar@{-}[d(0.5)] &
{\varphi_2,\varphi_2^c}\ar@{-}[d(0.5)] &
{\varphi_3,\varphi_3^c}\ar@{-}[d(0.5)] 
&
\\
{\cdots \quad}\ar@{->}[r]^{\displaystyle Q_{\Nm{1}}\quad}
&*+<0.4cm>[Fo]{SU(k)_0}\ar@{->}[r]^{\displaystyle Q_0}
&*+<0.4cm>[Fo]{SU(k)_1}\ar@{->}[r]^{\displaystyle Q_1}
&*+<0.4cm>[Fo]{SU(k)_2}\ar@{->}[r]^{\displaystyle Q_2}
&*+<0.4cm>[Fo]{SU(k)_3}\ar@{->}[r]^{\displaystyle Q_3}
&{\quad \cdots}
}
}
\end{equation*}
\caption{Segment of a quiver with $N$ nodes deconstructing a bulk $SU(k)$ gauge theory. To form a circular quiver consistent with an $S^1$ compactification, we identify $SU(k)_N = SU(k)_0$.}
\label{fig:quiver_circle}
\end{figure}

The deconstructed theory is represented as a quiver diagram, where each node represents a gauge group and the links represent bifundamental superfields. There can be additional matter content transforming under each node in the fundamental representation. In general, one may consider a variety of options for the link field variables. The link fields may simply be elementary chiral superfields, corresponding to a linear UV completion, or they may be fields in a non-linear sigma model corresponding to a more complicated dynamical UV completion. As long as supersymmetry is unbroken, this distinction is irrelevant for reproducing the leading physics of a supersymmetric extra dimension. As we will see, the choice of a non-linear sigma model for the link fields will be important when supersymmetry is broken. A segment of the standard deconstruction quiver is illustrated in Figure~\ref{fig:quiver_circle}, while the matter content is summarized in Table \ref{tab:matter}.

\begin{table}
\begin{center}
\begin{tabular}{c|ccccc}
& $SU(k)_0$ & $SU(k)_1$ & $SU(k)_2$ & $\cdots$ &  $SU(k)_{\Nm{1}}$ \T\B\\
 \hline 
 $Q_0$ & ${\Box}$ & $\overline\Box$ \T\B\\
 $Q_1$ & & ${\Box}$ & $\overline\Box$ \T\B\\
 $\vdots$ &&&& $\ddots$ \T\B\\
 $Q_{\Nm{1}}$ & $\overline\Box$ & & & &  $\Box$ \T\B\\
 \hline
 $\varphi_0,\varphi^c_0$ & $\Box,\overline\Box$ & & \T\B\\
 $\varphi_1,\varphi^c_1$ && $\Box,\overline\Box$ &  \T\B\\
 $\vdots$ &&&& $\ddots$ \T\B\\
 $\varphi_{\Nm{1}},\varphi^c_\Nm{1}$ &&&&& $\Box,\overline\Box$ \T\B\\
\end{tabular}
\end{center}
\caption{Matter content of an $N$-site deconstruction reproducing the physics of a 5D theory with bulk $SU(k)$ gauge symmetry. The $Q_i$ are link fields and the $\varphi_i, \varphi_i^c$ are vector-like matter fields reproducing bulk matter multiplets. \label{tab:matter}}
\end{table}

For simplicity, we will gently abuse notation by using the same labels $\varphi_i, Q_i$ for the superfields and their lowest scalar components. Let the $Q_i$ develop vacuum expectation values $v{\bf{1}}_{k\times k}$ due to other superpotential interactions in the UV which we will ignore for now\footnote{Choosing $\langle Q_i\rangle = v$ fixes the gauge choice except the diagonal one. Such a gauge choice is most convenient for calculations but will not affect physical conclusions}; in addition, we will assume that the vacuum moduli is already stabilized by these interactions. For simplicity -- and to reproduce the physics of a flat extra dimension -- we choose all vevs to be equal, but this requirement is not strictly necessary. Likewise, we impose a shift symmetry $i\rightarrow i+1$ such that the couplings $g_i$ are the same for each gauge group $SU(k)_{i}$. The mass terms for the gauge fields arising from this pattern of spontaneous symmetry breaking are then
\begin{eqnarray}
\mathcal{L} \ni \frac{1}{2}\sum_i g^2v^2 \left(A^{a\mu}_{i+1} - A^{a\mu}_{i}\right)^2
\label{eq:gauge_mass_term}
\end{eqnarray}
which give rise to a tower of states of mass
\begin{eqnarray}
m_n^2 = 2g^2v^2\sin^2 \left(\frac{\pi n}{N}\right)
\qquad n\in \left\{0,1,\cdots,N-1\right\}
\label{eq:gauge_mass_circle}
\end{eqnarray}
The correspondence with compactification of a 5D theory can be made clear by rewriting the masses in terms of the corresponding 5D quantities,
\begin{eqnarray}
R=\frac{N}{\sqrt{2}\pi gv} \quad\quad a=\frac{\sqrt{2}}{gv} 
\quad\quad m_n &=& \frac{2}{a} \sin\left(\frac{an}{2R}\right)
\quad\quad n \in (0, 1, ..., N-1)
\end{eqnarray}
where $R$ is the 5D radius in the continuum limit and $a$ is the lattice spacing for the apparent discretized extra dimensions. The tower of masses has a degeneracy when exchanging $n \leftrightarrow N-n$. These $N-n$ modes correspond to negative momenta states. In the limit $a/R\rightarrow 0$, the vector fields form uniform KK towers.  At this stage the pattern of higgsing is purely supersymmetric, so that the vector superfields and the $Q_i$ chiral superfields assemble themselves into a tower of $\mathcal{N}=2$ multiplets. One can verify this by checking that the $Q_i$ scalars obtain an identical KK mass spectrum from the $D$ terms:
\begin{align}
\mathcal{L}\ni \sum_i\frac{g^2_i}{2}D_i^a D_i^a \ni \frac{g^2}{2}
\sum_i \tr\left(Q_{i}^\dagger T^a Q_{i} - Q_{i-1}T^aQ_{i-1}^\dagger \right)^2
\label{eq:scalar_D_term}
\end{align}
Substituting $Q_i=Q^\dagger_i=v$ and expanding yields a mass term for the hermitian (traceless) part of $Q_i$, with the same form as Equation~\ref{eq:gauge_mass_term}. The trace of $Q_i$ remains massless but can be rendered massive by adding suitable interactions. Of the scalars, $2(k^2-1)$ real degrees of freedom form an $\mathcal{N}=2$ vector multiplet with the zero mode of the gauge bosons, and the rest are eaten to form a tower of massive $\mathcal{N}=2$ vector multiplets.

Thus far we have reproduced the spectra of an extra dimension compactified on $S^1$. Non-trivial boundary conditions such as the $S^1/\mathbb{Z}_2$ orbifold can be reproduced by breaking the circular quiver to form a linear quiver with distinct end-points, which we will explore in Sections \ref{sec:orbifold_gauge}-\ref{sec:orbifold_matter}.

\subsection{The Deconstructed Action}
\label{sec:deconstructed_action}
While the correspondence between mass spectra is appealing, to see the full reproduction of 5D physics, it is more illuminating to examine the correspondence at the level of the action. In this section, we demonstrate the deconstruction of the action explicitly. For simplicity, consider the deconstruction of a $U(1)$ gauge theory. Starting with a circular quiver, the action reads
\begin{eqnarray*}
S=\int d^4x \, \left( 
\int d^2 \theta \sum_i^N \frac{1}{4g_i^2} W^{\alpha i} W_{\alpha i} + \textrm{h.c.} +
\int d^4 \theta \sum_i^N Q_i^\dagger e^{V_i} Q_i e^{-V_{i+1}}
\right)
\end{eqnarray*}

As before, the $Q_i$ develop vevs and we can expand around the vacuum expectation value via $Q_i =v \left(1+\epsilon R\chi_i/\sqrt{2}\right)$, where $\chi_i \rightarrow 0$ at infinity, and $\epsilon,R$ are normalization factors that will be fixed by requiring agreement with the 5D action in the continuum limit. For the moment, the $Q_i$ transform linearly under the gauge group. As long as supersymmetry is unbroken, this choice is sufficient for reproducing 5D physics in the large $N$ limit. Expanding around the vev, the K\"{a}hler potential becomes
\begin{eqnarray}
\int d^4 \theta\, K(Q_i,Q_i^\dagger)&=&\int d^4 \theta \;  (v^\dagger v) \,\sum_i^N e^{-(V_{i+1}-V_{i})} 
\left[1+ \frac{\epsilon R}{\sqrt{2}}(\chi^\dagger_i + \chi_i)+\frac{\epsilon^2 R^2}{2} \chi_i^\dagger \chi_i\right] \notag \\
&=& \int d^4 \theta \;  (v^\dagger v) \sum_i^N \sum_n \frac{(-1)^n}{n!}(V_{i+1}-V_{i})^n \left[1+ \frac{\epsilon R}{\sqrt{2}}(\chi^\dagger_i + \chi_i)+\frac{\epsilon^2R^2}{2} \chi_i^\dagger \chi_i\right]
\end{eqnarray}

Let $\epsilon\, \delta V_i = V_{i+1} - V_{i}$. As we will see later, in the appropriate large $N$ limit, the masses for modes with large $\delta V_i$ will be high, so that $\delta V_i$ is of order unity. Using $\sum_i V_{i+1} - V_{i}=0$, we have

\begin{eqnarray}
\int d^4 \theta \, K(Q_i,Q_i^\dagger)=
\int d^4 \theta \;  &(v^\dagger v)&\sum_i^N \left\{1+\frac{\epsilon^2}{2} \left[ \delta V_i - \frac{R}{\sqrt{2}}(\chi^\dagger_i + \chi_i)\right]^2
+\frac{\epsilon R}{\sqrt{2}} (\chi^\dagger_i + \chi_i) - \right. \notag\\
&\quad&  \quad \left. \frac{\epsilon^2R^2}{4}(\chi_i^{\dagger 2}+\chi_i^{2})+ \mathcal{O}(\epsilon^3) \right\}
\label{eq:decon_nonabelian_spurious}
\end{eqnarray}
Since $v^\dagger v$ is a scalar, terms of the form $\int d^4\theta \chi$ and $\int d^4\theta \chi^2$ vanish. Rewriting all parameters

\begin{eqnarray}
\epsilon\rightarrow\frac{2\pi}{N} =\frac{a}{R}\qquad
\frac{1}{g^2} \rightarrow \frac{1}{g_4^2 N} \qquad
v^\dagger v \rightarrow \frac{2}{g_4^2 R^2 \epsilon^2}\cdot\frac{1}{N}
\label{eq:dictionary}
\end{eqnarray}
the Lagrangian takes the form
\begin{eqnarray}
\mathcal{L}&=&
\frac{1}{4g_4^2}\int d^2\! \theta \; \frac{1}{N}\!\sum_i^N  W^{\alpha i} W_{\alpha i} + \textrm{h.c.} + \notag \\
& &\quad \frac{1}{g_4^2 R^2}\int d^4 \!\theta\; \frac{1}{N}\!\sum_i^N \left[\delta V_i - \frac{R}{\sqrt{2}}(\chi^\dagger_i + \chi_i) \right]^2
 + \mathcal{O}(\epsilon)
\label{eq:decon_abelian_action}
\end{eqnarray}

We are now in a position to take the large-$N$ continuum limit, holding $R$ and $g_4$ fixed and taking the lattice spacing to zero, i.e. $\epsilon = a/R \rightarrow 0$. In this limit, $\frac{1}{N}\sum_i \rightarrow \int \frac{d\phi}{2\pi}$ and $\delta V_i= (V_{i+1}-V_{i})/\epsilon \rightarrow \partial_{\phi} V$. Since the action takes an analogous form to the full 5D action, we see that modes with high $\delta V_i$ have high masses, justifying the assumption that $\delta V_i$ is $\mathcal{O}(1)$. Then all the higher-order terms containing extra factors of $\epsilon$ can be ignored, and the action of a supersymmetric extra dimension is  fully recovered.

For non-abelian $SU(k)$ groups, the derivation becomes somewhat more complicated; we reserve the details for Appendix \ref{app:a}. The $\epsilon$ expansion in this case yields

\begin{eqnarray}
\int d^4 \theta \, K(Q_i, Q_i^\dagger)=\frac{2}{g_4^2 R^2} \int d^4 \theta \; 
\frac{1}{\epsilon^2 N}\sum_i^N \tr\, \left\{
\frac{\epsilon^2}{2} 
\left[f(L_{V_i})(\delta V_i) + \frac{R}{\sqrt{2}}\left(\chi^{\dagger}_i + e^{V_i}\chi_i^{\dagger}e^{-V_i}\right)\right]^2 \right. \notag \\
\left. + \frac{\epsilon R}{\sqrt{2}} \left(\chi^\dagger_i + \chi_i\right) -
\frac{\epsilon^2 R^2}{4} \left(\chi^{\dagger 2}_i + \chi_i^2\right) 
\right\} +\mathcal{O}(\epsilon^3)
\label{eq:decon_nonabelian_action}
\end{eqnarray}

where $L_{V_i}$ denotes the Lie derivative, i.e. $L_{V_i} A = [V_i, A]$, and $f(x) = (1-e^{x})/x$. Note the well-known formula $e^{A}\partial e^{-A}= f(L_A)\partial A$. The large-$N$ limit of Equation~\ref{eq:decon_nonabelian_action} reproduces Equation~\ref{eq:5D_nonabelian_action} as expected. The correspondence between 5D continuum and deconstruction for various parameters are shown in Table \ref{table:dictionary}.

\begin{table}[ht]
\begin{center}
\begin{tabular}{c|cccccccc}
5D continuum &
$R$ &
$g_4$ &
0 &
$\partial_\phi \Phi$&
$\int \frac{d\phi}{2\pi}$
\T\B\\\hline
Deconstruction &
$\frac{N}{\sqrt{2}\pi gv}$ &
$\frac{g}{\sqrt{N}}$ &
$\epsilon=\frac{a}{R}=\frac{2\pi}{N}$ &
$\frac{1}{\epsilon}(\Phi_{i+1}-\Phi_i)$&
$\frac{1}{N}\sum_i$
\T\B
\end{tabular}
\end{center}
\caption{Dictionary of deconstructed variables and their continuum counterparts.
\label{table:dictionary}}
\end{table}

\subsection{Adding matter}
To deconstruct additional matter, we add chiral superfields $(\varphi_i, \varphi^c_i)$ to each site transforming as $(\Box, \overline{\Box})$ of $SU(k)_i$. The Lagrangian for these fields is
\begin{eqnarray}
\mathcal{L}\ni \int d^4\theta \sum_{i} \tr(\varphi_i^\dagger e^{V_i}\varphi_i + \varphi_i^c e^{-V_i}\varphi^{c\dagger}_i ) +
\int d^2\theta \sum_i \tr(\lambda_i \varphi^c_i Q_i \varphi_{i+1} + m_i \varphi^c_i \varphi_i)
\end{eqnarray}
To take the continuum limit, we can substitute 
\begin{eqnarray}
\varphi_{i} = \Phi_i/\sqrt{N} \qquad
\varphi_{i+1}= (\Phi_i + \epsilon \, \delta \Phi_i)/\sqrt{N}
\qquad
M_i=\lambda_i v+ m_i \qquad
c_i=\frac{\sqrt{2}\lambda_i}{g}
\end{eqnarray}
The K\"{a}hler potential has no $\mathcal{O}(\epsilon)$ term and the continuum limit can be taken directly. As for the superpotential, we can expand the superpotential in $\epsilon$ as before to obtain
\begin{eqnarray}
W= \frac{1}{N}\sum_i \tr\left[(\lambda_i v+m_i)\Phi^c_i \Phi_{i} + 
 \lambda_i v\, \epsilon \left(\Phi^c_i \,\delta \Phi_i+\frac{R}{\sqrt{2}}\Phi^c_i\chi_i \Phi_i\right)\right] + \textrm{higher order terms}
\end{eqnarray}
Defining $M_i \equiv \lambda_i v + m_i$ and $c_i= \sqrt{2} \lambda_i/g$, we arrive at
\begin{eqnarray} 
W= \frac{1}{N}\sum_i \tr\left[M_i\Phi^c_i \Phi_{i} + 
 \frac{c_i}{R}\Phi^c_i\left( \delta \Phi_i+\frac{R}{\sqrt{2}}\chi_i \Phi_i \right)\right] + \mathcal{O}(\epsilon)
\label{eq:matter_mass}
\end{eqnarray}
We see that in order to recover a proper extra dimension, the $M_i, c_i$ must be tuned to their 5D counterpart, with $M_i=M$ the 5D mass, and $c_i=1$ in the large $N$ limit. For our discussion, we will fix $\lambda_i = \sqrt{2} g$ and $m_i = \sqrt{2}gv$, which will lead to exact formulae for the masses. The superfields $(\Phi_i, \Phi^c_i)$ then form a tower of $\mathcal{N}=2$ hypermultiplets, and their masses are given by
\begin{eqnarray} 
m_n^2 &=& \left(\frac{2}{a}\right)^2 \sin^2\left(\frac{an}{2R}\right)
\qquad n\in (0, 1,..., N-1)
\label{eq:matter_mass_tower}
\end{eqnarray}
Again, for a circular quiver, there is a degeneracy $n\leftrightarrow N-n$, corresponding to positive and negative momenta modes. As the lattice spacing is taken to zero, the 5D action and the bulk matter spectrum is recovered.

\subsection{Orbifold (gauge)}
\label{sec:orbifold_gauge}
\begin{figure}[htb!]
\vspace{0.25cm}
\begin{align*}
\resizebox{11cm}{!}{
\xymatrix{
*+<0.4cm>[Fo]{SU(k)_{0 \,} }\ar@{->}[r]^{\displaystyle Q_0}
&*+<0.4cm>[Fo]{SU(k)_{1\,}}\ar@{->}[r]^{\displaystyle Q_1}
&*+<0.4cm>[Fo]{SU(k)_{2\,}}\ar@{->}[r]^{\displaystyle Q_2}
&{\quad \cdots \quad}\ar@{->}[r]^{\!\!\!\! \displaystyle Q_{\Nm{2}}}
&*+<0.20cm>[Fo]{SU(k)_{\Nm{1}}  }
}
}
\end{align*}
\caption{A quiver with $N$ nodes deconstructing an $SU(k)$ gauge theory on an $S^1/\mathbb{Z}_2$ orbifold. There is one more gauge node than link field, resulting in $\mathcal{N}=1$ supersymmetry for the massless vector mode.}
\label{fig:quiver_orbifold}
\end{figure}
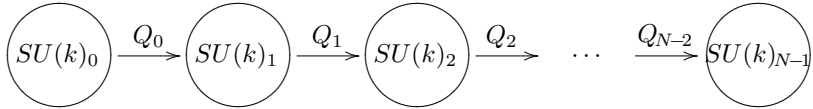

So far we have only considered deconstruction of a circular quiver; in order to produce phenomenologically viable theories, it is important to remove part of the $\mathcal{N}=2$ supersymmetry in the low energy 4D theory. For the vector multiplet, the zero mode for the adjoint scalars needs to be removed. This can be accomplished by putting the extra dimension on a $S^1/\mathbb{Z}_2$ orbifold. By imposing different parity on different components of the same multiplet at the orbifold fixed point, $\mathcal{N}=2$ supersymmetry can be broken. Equivalently, one can view the orbifold as an interval $(0,\pi R)$; assigning orbifold parity is equivalent to assigning Dirichlet/Neumann boundary conditions at $0$ or $\pi R$. In terms of quivers, Dirichlet boundaries for the $Q_i$ can be achieved by removing $Q_{\Nm{1}}$, effectively forcing it to vanish. The quiver diagram is shown in Figure~\ref{fig:quiver_orbifold}. The link fields produce gauge boson masses of the form $m^2_A \sim \sum_i (A_{i+1} - A_i)^2$, force the massless vector to satisfy $A_{i+1}^{(0)}=A_{i}^{(0)}$, effectively resulting in a Neumann boundary condition. The deconstruction of Dirichlet/Neumann boundary conditions can be generalized to the entire tower (see Appendix~\ref{app:b} for a full derivation).

The existence of a chiral massless mode is robust against variations of the gauge couplings and link-field vevs. With one less $Q_i$, the massless vector multiplet no longer has a partner chiral field, and the $\mathcal{N}=2$ supersymmetry of the zero mode is reduced to $\mathcal{N}=1$. One might worry that the gauge groups at the end-point are anomalous; additional spectator matter fields can be added to remove the anomaly, which will be discussed in Section~\ref{sec:anomalies}.

To compute masses for the $Q_i$ fields in the linear quiver, one considers all the $D$ terms
\begin{eqnarray}
\mathcal{L}\ni \frac{g^2}{2}
\left[
\sum_{i=1}^{N-2} \tr\left(Q_{i}^\dagger T^a Q_{i} - Q_{i-1}T^aQ_{i-1}^\dagger \right)^2 +
\tr\left(Q_{\Nm{2}} T^a Q_{\Nm{2}}^\dagger\right)^2 +
\tr\left(Q_{0}^\dagger T^a Q_{0}\right)^2
\right]
\label{eq:scalar_D_term}
\end{eqnarray}
Expanding the $Q_i$ around their vacuum expectation values, the masses for the hermitian traceless parts of the $Q_i$ are
\begin{eqnarray}
m_n^2=2g^2v^2\sin^2\left( \frac{\pi n}{2N} \right) =
\left(\frac{2}{a}\right)^2 \sin^2\left(\frac{an}{2R}\right)
\qquad
n\in \left\{ 1,\cdots,N-1\right\}
\label{eq:gauge_mass_orbifold}
\end{eqnarray}
There is an extra factor of two when comparing Equation~\ref{eq:gauge_mass_orbifold} to Equation~\ref{eq:gauge_mass_circle}, and $R$ is a factor of two larger compared to before. This is expected since the quiver now corresponds to the segment $(0, \pi R)$ instead of $(0, 2\pi R)$. The extra factor also breaks the degeneracy $n\rightarrow N-n$, as expected with an orbifold. The continuum limit then only contains half of the full KK tower. The new dictionary for the parameters is then
\begin{eqnarray}
R=\frac{\sqrt{2}N}{\pi gv} \qquad a=\frac{\sqrt{2}}{gv} \qquad \epsilon=\frac{a}{R}=\frac{\pi}{N}
\qquad
m^2_n=\left(\frac{2}{a}\right)^2 \sin^2\left(\frac{an}{2R}\right)
\end{eqnarray}
For the gauge fields, the mass terms are
\begin{eqnarray}
\mathcal{L} \ni \frac{1}{2}\sum_{i=0}^{N-2} g^2v^2 \left(A^{a\mu}_{i+1} - A^{a\mu}_{i}\right)^2
\end{eqnarray}
Where the term $(A^\mu_{\Nm{1}} - A^\mu_0)^2$ is now absent, this leads to a tower of masses
\begin{eqnarray}
m_n^2=2g^2v^2\sin^2\left( \frac{\pi n}{2N} \right) =\left(\frac{2}{a}\right)^2 \sin^2\left(\frac{an}{2R}\right)
\qquad
n\in \left\{ 0,1,\cdots,N-1\right\}
\end{eqnarray}
The massless mode for the gauge bosons is retained. All massive gauge bosons pair up with the massive tower of $Q_i$ to form $\mathcal{N}=2$ multiplets, and the desired physics of the orbifold is recovered. 

Note that we have so far discussed the case of $SU(k)$ nodes, for which the individual link field vevs are $D$-flat and there is no complication in taking a linear quiver. This is no longer the case for abelian $U(1)$ nodes, where the link field vevs cease to be $D$-flat at the endpoints of the linear quiver. However, this can be dealt with by adding Fayet-Iliopoulos terms at the endpoints as discussed in \cite{Dudas:2003iq}. 

\subsection{Orbifold (matter)}
\label{sec:orbifold_matter}
\begin{figure}[htb!]
\begin{equation*}
\resizebox{12cm}{!}{
\xymatrix{
{\varphi_0,\varphi_0^c}\ar@{-}[d(0.5)] &
{\varphi_1,\varphi_1^c}\ar@{-}[d(0.5)] &&
{\varphi_{\Nm{2}},\varphi_{\Nm{2}}^c}\ar@{-}[d(0.5)] &
{\varphi_{\Nm{1}}}\ar@{-}[d(0.45)] 
\\
*+<0.4cm>[Fo]{SU(k)_0}\ar@{->}[r]^{\displaystyle Q_0}
&*+<0.4cm>[Fo]{SU(k)_1}\ar@{->}[r]^{\displaystyle Q_1}
&{\quad \cdots \quad}\ar@{->}[r]^{\! \displaystyle Q_{\Nm{3}}}
&*+<0.05cm>[Fo]{SU(k)_{\Nm{2}}}\ar@{->}[r]^{\displaystyle \! Q_{\Nm{2}} }
&*+<0.05cm>[Fo]{SU(k)_{\Nm{1}} }
}
}
\end{equation*}
\caption{A quiver with $N$ nodes deconstructing an $SU(k)$ gauge theory on an $S^1/\mathbb{Z}_2$ orbifold, including chiral matter. There is one more $\varphi_i$ than $\varphi^c_i$, resulting in an extra chiral $\varphi_i$ with $\mathcal{N}=1$ supersymmetry.}
\label{fig:quiver_orbifold_matter}
\end{figure}
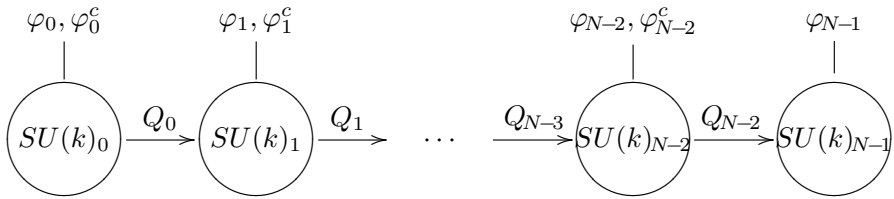

Besides gauge bosons, a phenomenologically viable model also requires chiral matter. Analogous to the vector multiplet case in Section \ref{sec:orbifold_gauge}, chiral matter can be obtained by removing an appropriate matter field at one of the end-points, corresponding to forcing a Dirichlet boundary condition and Neumann boundary condition for the conjugate field. We leave the detailed mass matrix calculation in Appendix~\ref{app:b}. For example, to get a chiral field in the fundamental representation, one can remove $\varphi^c_{\Nm{1}}$, so that an extra $\varphi$ field will remain unpaired in the low energy theory, leaving $\mathcal{N}=1$ supersymmetry at the level of the zero modes.  The corresponding quiver diagram is shown in Figure~\ref{fig:quiver_orbifold_matter}.

As before, the superpotential remains the same except all couplings with $\varphi^c_{\Nm{1}}$ are removed. The masses for the matter fields are

\begin{eqnarray}
m_{\varphi, n}^2 &=& \left(\frac{2}{a}\right)^2 \sin^2\left(\frac{an}{2R}\right) 
\qquad n \in (0, 1, \cdots, N-1)
\notag \\
m_{\varphi^c, n}^2 &=& \left(\frac{2}{a}\right)^2 \sin^2\left(\frac{an}{2R}\right)
\qquad n \in (1, \cdots, N-1)
\end{eqnarray}
As in the case of the circular quiver, in order to produce a matter spectrum with an appropriate 5D limit, the couplings between the $\varphi_i$ must be tuned. However, the existence of a massless chiral field is guaranteed even without tuning the parameters, as the extra $\varphi_i$ has no partner to pair up in order to become massive. 

\subsection{Anomalies}
\label{sec:anomalies}

In general the end-point gauge groups of the deconstruction are anomalous, whereas the ``bulk'' gauge groups are anomaly-free since all link and matter fields are vector-like on these sites. The gauge orbifold leaves the $SU(k)_0$ and $SU(k)_{N-1}$ gauge groups with $k$ surplus fundamentals and anti-fundamentals, respectively. For the matter, a fundamental chiral zero mode leaves the $SU(k)_{N-1}$ gauge group with an extra fundamental, while an anti-fundamental zero mode leaves the $SU(k)_{0}$ group with an extra anti-fundamental. The gauge orbifold anomaly may be addressed by adding $k$ anti-fundamental spectator fields $S_0^{c,a}$ ($a=1,\dots k$) to the $SU(k)_0$ group and $k$ fundamental spectator fields $S_{N-1}^a$ to the $SU(k)_{N-1}$ group. These spectator fields do not acquire a vev, but may be rendered massive by irrelevant superpotential terms of the form $\sum_a^{k}S^{c,a}_{0} Q_1 \cdots Q_{N-1} S^{a}_{\Nm{1}}$. Strictly speaking, these spectators are massless in the $\mathcal{N} \to \infty$ limit, but are massive in a finite quiver. Taken on their own, the anomalies from matter orbifolds are somewhat thornier; while we could add suitable spectator fields in analogy with the gauge orbifold case, these spectators cannot be rendered massive by an analogous irrelevant superpotential due to their gauge representations. It may be the case that the anomaly is cancelled by the full set of matter representations in a complete quiver, but if not, it is typically still the case that the matter orbifold anomaly only partially cancels the gauge orbifold anomaly, and the remaining gauge orbifold anomaly may be cancelled as above with a reduced number of spectator fields $S^a_{N-1}$ and  $S^{c,a}_0$.

\section{Radion Mediation} \label{sec:radion}

All the quivers considered above have unbroken $\mathcal{N}=1$ supersymmetry in the zero mode spectrum. In this section, we proceed to deconstruct Scherk-Schwarz supersymmetry breaking by exploiting its relationship to radion mediation. To deconstruct radion mediation, we introduce a 4D analogue of the radion and radion-matter couplings. The natural proposal is to introduce a chiral superfield normalized analogously to the radion,
\begin{eqnarray}
T=\frac{1}{R}\left(R+ iB_{5} + \theta \Psi^{5} + \theta^2 F_{T}\right)
\end{eqnarray}
To reproduce the physics of an extra dimension, we take a universal $T$ without any site dependence; for more general cases, a $T_{i}$ for each site can be introduced. 

Since deconstruction reproduces 5D physics at the level of the action, it is tempting to directly introduce analogous radion-matter couplings to the deconstructed action according to Equation~\ref{eq:radion_replacement}. However, introducing radion-like couplings to the linear realization of the supersymmetric deconstruction does not reproduce all the physics of radion mediation, as terms of the form $\int d^4\theta T^\dagger \chi_i \chi_i$ required for the majorana gaugino mass spectrum are absent. One potential remedy is to add higher-dimensional operators proportional to $Q^\dagger_i Q^\dagger_i$, but additional matter needs to be introduced to render such a term gauge invariant. More problematically, naively coupling $T$ universally to matter terms in the action can potentially give rise to spurious terms such as $\frac{1}{\epsilon}\int d^4 \theta T^\dagger \chi_i$, leading to divergent behavior in the large-$N$ limit.

The remedy is to promote the link field variables $Q_i$ into  $SL(k,\mathbb{C})$-valued non-linear sigma fields, and expand around the vacuum expectation value
\footnote{For a $U(1)$ theory, writing $Q_i=v\exp(\chi_i/gv)$ suffices as the spurious terms $\int d^2\theta \sum_i(\chi^\dagger_i + \chi_i)$ become gauge invariant and can be removed by hand directly.  }
\begin{align}
Q_i= v \exp\left(\frac{\chi_i}{gv} \right) = v\exp\left(\frac{\epsilon R\chi_i}{\sqrt{2}}\right)
\label{eq:nonlinear_sigma}
\end{align}
This is not surprising; while it was perfectly adequate to take the $Q_i$ as linear fields in the supersymmetric limit, the exact reproduction of the full 5D physics on a lattice requires Wilson loop variables, which are essentially non-linear sigma fields. As such, it implies that the theory is not UV complete in itself; the deconstruction grows strongly coupled around $4 \pi v$ and requires a further UV completion, albeit one that can be readily achieved in four dimensions.

  In terms of these $\chi_i$ fields, the K\"{a}hler potential takes the form
\begin{eqnarray}
\int d^4 \theta \sum_i^N  \tr\, v^\dagger v \, e^{\frac{\epsilon R\chi^\dagger_i}{\sqrt{2}}} e^{V_i} e^{\frac{\epsilon R\chi_i}{\sqrt{2}}} e^{-V_{i+1}}
\end{eqnarray}

Expanding the exponential and taking the large $N$ limit, we find that working in terms of non-linear sigma variables introduces one additional quadratic term relative to Equation~\ref{eq:decon_nonabelian_spurious}:
\begin{eqnarray*}
\mathcal{L} \supset \frac{1}{g_4^2 R^2 \epsilon^2}\int d^4 \!\theta\; \frac{1}{N}\!\sum_i^N\frac{\epsilon^2R^2}{4}(\chi^{\dagger 2}+\chi^{2})
\label{eq:newterm}
\end{eqnarray*}

This new term comes from the second order term in the exponential, and it exactly cancels the spurious $\chi^{\dagger 2}+\chi^2$ term already appearing in Equation~\ref{eq:decon_nonabelian_spurious}. In addition, by demanding the $Q_i$ to be $SL(k,\mathbb{C})$ valued, terms proportional to $\tr (\chi + \chi^\dagger)$ vanish. Now that all the spurious terms are canceled, the radion coupling can be added as prescribed in Equation \ref{eq:radion_replacement}, more explicitly,
\begin{eqnarray}
\int d^4 \!\theta\; \sum_i Q_i^\dagger e^{V_i} Q_i e^{-V_{i+1}} &\longrightarrow&
\int d^4 \!\theta\; \frac{2}{T+T^\dagger} \sum_i^N Q_i^\dagger e^{V_i} Q_i e^{-V_{i+1}} \notag \\
\int d^4 \!\theta\; \sum_i\varphi_i^\dagger e^V \varphi_i &\longrightarrow&
\int d^4 \!\theta\; \frac{T+T^\dagger}{2}\, \sum_i\varphi_i^\dagger e^V \varphi_i \notag \\
\int d^4 \!\theta\; \sum_i \varphi_i^c e^{-V} \varphi_i^{c\dagger} &\longrightarrow&
\int d^4 \!\theta\; \frac{T+T^\dagger}{2}\, \sum_i\varphi_i^c e^{-V} \varphi_i^{c\dagger} \notag \\
\frac{1}{2g^2}\int d^2\!\theta \, \sum_i\tr\, W_i^\alpha W_{i\alpha} &\longrightarrow&
\frac{1}{2g^2}\int d^2\!\theta \, T \sum_i\tr\,  W^\alpha_i W_{i\alpha}
\label{eq:radion_replacement_deconstruction}
\end{eqnarray}
and the 5D action with radion coupling is fully reproduced. Thus we have encountered an interesting lesson: although the physics of a supersymmetric extra dimension can be reproduced equally well using linear fields or non-linear sigma fields for the link variables, once supersymmetry is broken we require the full non-linear sigma model. 

\subsection{The mass spectrum}
By incorporating a universal ``radion'' in our quiver, we can obtain a full deconstruction of the supersymmetry breaking Scherk-Schwarz mechanism. Here we summarize the findings and discuss the relevant physics; the detailed computations of the spectrum are left to Appendix~\ref{app:b}. 

The radion couplings only affect the masses for gauginos and matter-field scalars. With the identification $F_T=2\alpha$ for a circular quiver, the gaugino masses and eigenstates are 
\begin{align}
\bm{\lambda}_{nj}^{\pm} = \frac{e^{ip_n aj}}{\sqrt{2}}\binom{e^{-\frac{ip_n a}{2}}}{\pm i} 
\qquad p_n a=\frac{2\pi n}{N}&= \frac{an}{R} 
\qquad
n \in (0, 1, ..., N-1) 
\label{eq:app_eigenstates}
\\
m_{n}^{\pm} = \left(\frac{2}{a}\right) \left| \sin \left( \frac{p_n a}{2} \right)\pm \frac{a F_T}{4R} \right| &= \left(\frac{2}{a}\right) \left| \sin \left( \frac{n a}{2R} \right)\pm \frac{a\alpha}{2R} \right| 
\end{align}
We see that there are two towers of majorana fermions, with a uniform mass splitting of $\pm \alpha/R$. The $n \rightarrow N-n$ degeneracy remains, and in the large $N$ limit, one obtains the mass spectrum $|n\pm \alpha|/R$. For the matter fields, the tree level mass spectrum and eigenstates are exactly identical.

Less trivially, we can also combine radion mediation with the deconstruction of a $S^1/\mathbb{Z}_2$ orbifold. The quiver diagram is again the one shown in Figure~\ref{fig:quiver_orbifold_matter}, now with the addition of radion couplings. The identification between the radion expectation value and Scherk-Schwarz twist is $F_T = \alpha$ for the orbifold. The gaugino masses and eigenstates in this case are
\begin{align}
\bm{\lambda}_{nj}^{\pm}= \frac{1}{\sqrt{N}}
\binom{\cos\left[p_n a(j+\sfrac{1}{2})\right]}{\pm\sin[p_n a(j+1)]}
\qquad p_n a=\frac{\pi n}{N}&= \frac{an}{R} 
\qquad
n \in (0, 1, ..., N-1) 
\label{eq:app_eigenstates}
\\
m_{n}^{\pm} = \left(\frac{2}{a}\right) \left| \sin \left( \frac{p_n a}{2} \right)\pm \frac{a F_T}{2R} \right| &= \left(\frac{2}{a}\right) \left| \sin \left( \frac{n a}{2R} \right)\pm \frac{a\alpha}{2R} \right| 
\end{align}

We see that in the continuum limit the $aj$ factor becomes $x_5$, and the $\lambda_{n}$ spinor goes like $\sim \cos(p_n x_5)$ which has Neumann boundary conditions at both of the orbifold fixed points $0$ and $\pi R$, whereas the $(-i\psi_{\chi})_{n}$ component goes like $\sim \sin(p_nx_5)$ and has Dirichlet boundary conditions. The $n=0$ mode is unique, since $\psi_{\chi_j}=0$ in this case. This is unsurprising since we expect the quiver to produce an $N=1$ vector multiplet with a supersymmetry breaking gaugino mass term in the low energy theory. As for the scalars of the hypermultiplet, the masses are the same as the gauginos but the eigenstates depend on which chiral superfield is removed. Writing the hypermultiplet as $\bm{\varphi}_j=(\varphi_j, \varphi^{c\dagger}_j)$, if $\varphi_{N-1}^c$ is removed, the eigenstates and masses will be
\begin{align}
\bm{\varphi}_{nj}^{\pm}= \frac{1}{\sqrt{N}}
\binom{\cos\left[p_n a(j+\sfrac{1}{2})\right]}{\pm\sin\left[p_n a(j+1)\right]}
\qquad p_n a=\frac{\pi n}{N}&= \frac{an}{R} 
\qquad
n \in (0, 1, ..., N-1) 
\label{eq:app_eigenstates}
\\
m_{n}^{\pm} = \left(\frac{2}{a}\right) \left| \sin \left( \frac{p_n a}{2} \right)\pm \frac{a F_T}{2R} \right| &= \left(\frac{2}{a}\right) \left| \sin \left( \frac{n a}{2R} \right)\pm \frac{a\alpha}{2R} \right| 
\end{align}
Again, the mass spectrum reproduces the boundary conditions in the continuum limit with $\varphi^c$ and $\varphi$ having Dirichlet and Neumann boundary conditions, respectively.

\section{Examples} \label{sec:ex}

Having accomplished our objective of reproducing the spectrum of Scherk-Schwarz supersymmetry breaking in four dimensions, we can now collect the tools developed in previous sections and apply them to a few illustrative examples. In Section ~\ref{subsec:mssm}, we present a toy model with a deconstructed bulk $SU(2)$ gauge group and bulk Higgs fields. In Section ~\ref{subsec:lightscalar}, we present possible ways to deconstruct an inverted spectrum where supersymmetry breaking leads to heavy fermions and light bosons. In Section ~\ref{subsec:folded}, we deconstruct a simplified model of folded supersymmetry and comment on potential natural models relevant to phenomenology.

\subsection{Bulk Higgs}
\label{subsec:mssm}
In this section we present a toy model with an $SU(2)$ bulk gauge group and bulk Higgs fields analogous to the model presented in \cite{Barbieri:2001yz}. We would like to reproduce the physics of an extra dimension with zero modes for both up-type and down-type Higgs doublets. The quiver is as follows:

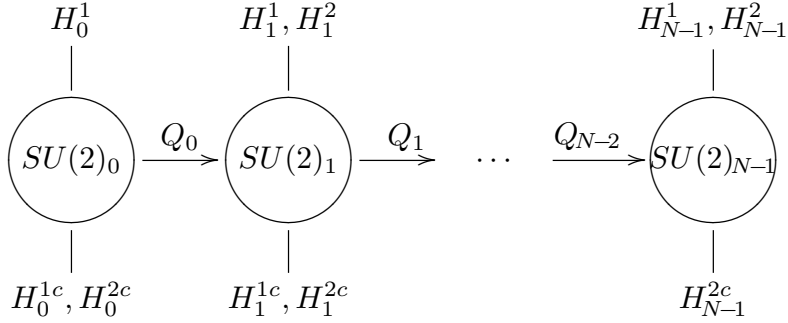
\begin{figure}[htb]
\begin{eqnarray*}
\resizebox{12cm}{!}{
\xymatrix{
{H^1_0}\ar@{-}[d(0.5)] &
{H^1_1, H^2_1}\ar@{-}[d(0.5)] &&
{H^1_{\Nm{1}}, H^2_{\Nm{1}}}\ar@{-}[d(0.5)]
&
\\
*+<0.47cm>[Fo]{SU(2)_0}\ar@{->}[r]^{\displaystyle Q_0}
&*+<0.47cm>[Fo]{SU(2)_1}\ar@{->}[r]^{\displaystyle \quad Q_1}
&{\quad \cdots \quad}\ar@{->}[r]^{\displaystyle Q_{\Nm{2}} \quad}
&*+<0.1cm>[Fo]{SU(2)_{\Nm{1}}}\\
{H^{1c}_0, H^{2c}_0}\ar@{-}[u(0.5)] &
{H^{1c}_1, H^{2c}_1}\ar@{-}[u(0.5)] &&
{H^{2c}_{\Nm{1}}}\ar@{-}[u(0.5)]
&
}
}
\end{eqnarray*}
\caption{Deconstruction of two Higgs fields living in a 5D bulk. The low energy theory contains two chiral Higgs fields $H_u, H_d$, with small SUSY breaking radion couplings added ($\alpha \ll 1$).}
\end{figure} 

Note that the conventions of our deconstruction require us to remove an $H^{1c}$ chiral superfield at one end-point of the quiver and an $H^2$ chiral superfield at the other end-point in order to obtain chiral zero modes, due to the choice of representations for the link fields $Q_i$. Consequently, the fields $H^{1c}_{\Nm{1}}$ and $H^2_{0}$ are absent in the quiver. The removal of chiral fields will generally make the gauge nodes at the end-points anomalous. We can cancel the anomalies by adding spectators chiral fields as discussed in Section ~\ref{sec:anomalies}. Alternately, we can assume the anomalies are accounted for by additional physics at the cutoff of the link field non-linear sigma model.

Before turning on couplings to the radion superfield, the K\"{a}hler potential is given by
\begin{equation}
K=\sum_{i=0}^{N-1} \left[ H_i^{1\dagger} e^{V_i} H_i^1 + H_i^{2c} e^{-V_i} H_i^{2c\dagger}\right]
+ \sum_{i=1}^{N-1} H_i^{2\dagger} e^{V_i} H_i^2 
+ \sum_{i=0}^{N-2} H_i^{1c} e^{-V_i} H_i^{1c\dagger} + K(Q_i)
\end{equation}

The corresponding superpotential is
\begin{equation}
W=\frac{g}{\sqrt{2}}\left[
\sum_{i=0}^{N-2}\tr( H^{1c}_i Q_i H^1_{i+1} + H^{2c}_i Q_i H^2_{i+1}) -
\sum_{i=0}^{N-2}\tr( v H^{1c}_i H^1_i) - \sum_{i=0}^{N-1} \tr(vH^{2c}_i H^2_i)
\right]
+ W_{\textrm{gauge}}
\end{equation}
Given the chiral endpoints of the quiver, the massless chiral modes can be identified with $H_u, H_d$:
\begin{eqnarray}
H_u = \frac{1}{\sqrt{N}} \sum_{i = 0}^{N-1} H_i^1 \qquad
H_d = \frac{1}{\sqrt{N}} \sum_{i = 0}^{N-1} H_i^{2c}
\label{eq:hu_hd}
\end{eqnarray}
In order to induce couplings of the form $H_u H_d$, we can add terms to the superpotential
\begin{eqnarray}
W \ni  \frac{\gamma}{R}\frac{N}{N-2} \sum_{i=1}^{N-2} H_i^1 H_i^{2c}
\end{eqnarray}
We assume these terms are identical at each site to reproduce the 5D spectrum, but in general these couplings may vary from site to site.

We now break supersymmetry by adding the usual radion mediation terms in accordance with Equation~\ref{eq:radion_replacement_deconstruction} with the identification $F_T=\alpha$ for an orbifold. In general, the mass eigenstates of the Higgs fields are complicated functions of $\gamma$ and $\alpha$. However, if we assume that $\gamma,\alpha \ll 1$, the eigenstates in Equation~\ref{eq:hu_hd} are good approximations at leading order. In this limit, the relevant interactions for the zero mode degrees of freedom in the Higgs sectors are
\begin{eqnarray}
\mathcal{L}\ni 
\int d^4 \theta \;\left(1+ \frac{\alpha}{R}\theta^2 +\frac{\alpha}{R}\theta^{\dagger 2}\right)\tr \left( H_u^\dagger  e^{V} H_u + H_d\, e^{-V} H_d^\dagger \right)
+ \int d^2\theta \, \frac{\gamma}{R}H_u H_d
\end{eqnarray}
accompanied by the expected towers of massive states. The resulting mass terms for the zero mode scalar Higgs doublets are
\begin{eqnarray}
\frac{\alpha^2 + \gamma^2}{R^2}\left( |H_u|^2 - |H_d|^2 \right) + 
\frac{2\alpha \gamma}{R^2}H_u H_d + \textrm{h.c.}
\end{eqnarray}
There are also parametrically light higgsino modes with masses $\gamma/R$.  Thus we reproduce the results of \cite{Barbieri:2001yz}.

\subsection{Orbifold with light scalars}
\label{subsec:lightscalar}
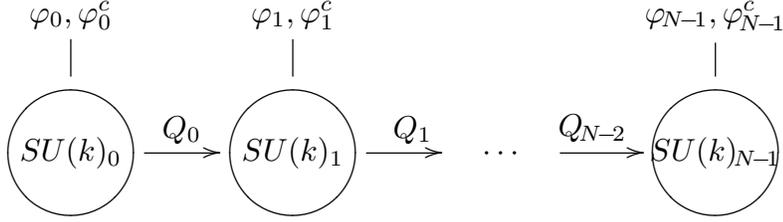
\begin{figure}[htb]
\begin{eqnarray*}
\resizebox{12cm}{!}{
\xymatrix{
{\varphi_0,\varphi_0^c}\ar@{-}[d(0.45)] &
{\varphi_1,\varphi_1^c}\ar@{-}[d(0.45)] &&
{\varphi_{\Nm{1}},\varphi_{\Nm{1}}^c}\ar@{-}[d(0.45)]
&
\\
*+<0.5cm>[Fo]{SU(k)_0}\ar@{->}[r]^{\displaystyle Q_0}
&*+<0.5cm>[Fo]{SU(k)_1}\ar@{->}[r]^{\displaystyle \quad Q_1}
&{\quad \cdots \quad}\ar@{->}[r]^{\displaystyle Q_{\Nm{2}} \quad}
&*+<0.15cm>[Fo]{SU(k)_{\Nm{1}}}
}
}
\end{eqnarray*}
\caption{An orbifold quiver with massless vector and massive matter content. Light scalars are obtained by tuning the supersymmetry breaking radion coupling.}
\label{fig:quiver_lightscalar}
\end{figure}

In the previous subsection, we considered an example where both scalar and fermion zero modes are parametrically light due to the smallness of SUSY-breaking terms relative to the separation of higher modes. As a second example, we consider possible ways to produce light scalars and heavy fermions in a deconstructed orbifold. A spectrum of light scalars and heavy fermions is crucial for folded supersymmetry and other phenomenological applications of 5D orbifolds. To proceed, we start with the orbifold quiver in Figure~\ref{fig:quiver_lightscalar}.

With the standard action, there will be a massless vector multiplet, but all the matter will be massive as they pair up to acquire Dirac masses. One option for generating light matter scalars, in the spirit of radion mediation, is to introduce a universal radion coupling in accordance with Equation~\ref{eq:radion_replacement_deconstruction}. The computation of the masses is described in Appendix ~\ref{app:b}; the resulting masses and eigenstates are

\begin{align}
\bm{\varphi}_{nj}^{\pm}= \frac{1}{\sqrt{N}}
\binom{\cos\left[p_na(j+\sfrac{1}{2})\right]}{\mp\sin\left[p_na(j+1)\right]}
&\qquad p_na=\frac{\pi (2n+1)}{2N+1}= \frac{a\left(n +\sfrac{1}{2}\right)}{R  \left(1+\sfrac{1}{N}\right)}
\qquad
n \in (0, 1, ..., N-1) 
\notag\\
m_{n}^{\pm} &= \left(\frac{2}{a}\right) \left| \sin \left( \frac{p_na}{2} \right)\pm \frac{a F_T}{2R} \right| 
\end{align}
To make the lightest scalar massless at tree level, we must fix
\begin{align}
F_T=\left(\frac{2R}{a}\right)\sin \left[ \frac{a}{4R\left(1+\sfrac{1}{N}\right)} \right]=
\left(\frac{2N}{\pi}\right)\sin \left[ \frac{\pi}{4\left(N+1\right)} \right]
\end{align}
which unsurprisingly corresponds to $F_T = \alpha = 1/2$ in the large $N$ limit. In general, there are radiative corrections to the masses, and the scalars that are massless at tree level acquire mass at one loop. 

%
%

\subsection{Folded supersymmetry}
\label{subsec:folded}
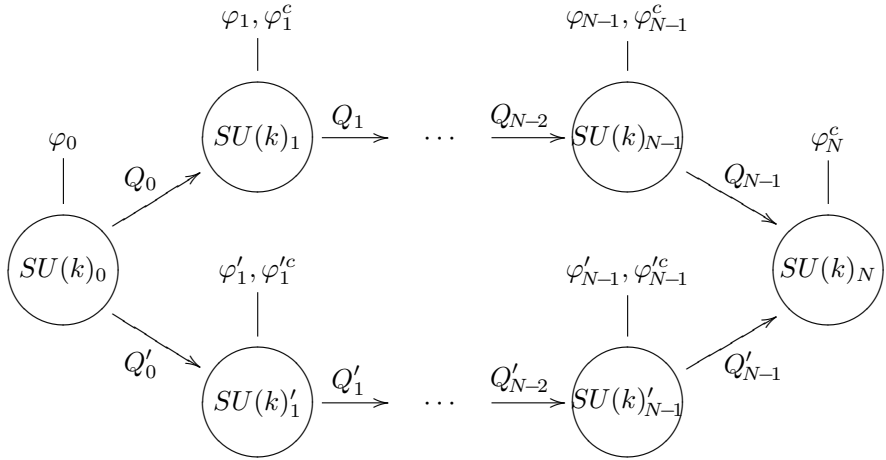
\begin{figure}[htb]
\begin{eqnarray*}
\resizebox{12cm}{!}{
\xymatrix{
 &
{\varphi_1,\varphi_1^c}\ar@{-}[d(0.45)] &&
{\varphi_{\Nm{1}},\varphi_{\Nm{1}}^c}\ar@{-}[d(0.45)]
& 
\\
{\varphi_0}\ar@{-}[d(0.5)]
&*+<0.5cm>[Fo]{SU(k)_1}\ar@{->}[r]^{\displaystyle Q_1}
&{\quad \cdots \quad}\ar@{->}[r]^{\displaystyle Q_{\Nm{2}} \quad}
&*+<0.15cm>[Fo]{SU(k)_{\Nm{1}}}
&
{\varphi_{\!N}^c}\ar@{-}[d(0.5)]
\\
*+<0.5cm>[Fo]{SU(k)_0}
\ar@{ ->}+<18pt,16pt>;[ru(0.6)]^{\displaystyle Q_0}
\ar@{->}+<18pt,-16pt>;[rd(0.6)]_{\displaystyle Q'_0}
 &
{\varphi'_1,\varphi_1^{\prime c}}\ar@{-}[d(0.5)] &&
{\varphi'_{\Nm{1}},\varphi_{\Nm{1}}^{\prime c}}\ar@{-}[d(0.5)]
&
*+<0.4cm>[Fo]{SU(k)_N}
\ar@{<-}+<-19pt,17pt>;[lu(0.6)]_{\displaystyle Q_{\Nm{1}}}
\ar@{<-}+<-19pt,-17pt>;[ld(0.6)]^{\displaystyle Q'_{\Nm{1}}}
\\
&*+<0.5cm>[Fo]{SU(k)'_1}\ar@{->}[r]^{\displaystyle Q'_1}
&{\quad \cdots \quad}\ar@{->}[r]^{\displaystyle Q'_{\Nm{2}} \quad}
&*+<0.15cm>[Fo]{SU(k)'_{\Nm{1}}}
}
}
\end{eqnarray*}
\caption{A ``folded'' quiver with a $\mathbb{Z}_2$ bifold symmetry. The bifold symmetry exchanges the primed gauge groups and matter with the unprimed fields. In order to decouple the two quivers, the $\mathbb{Z}_2$ symmetry is spontaneously broken by $\langle Q_0'\rangle = \langle Q_{\Nm{1}}'\rangle = 0$}
\label{fig:quiver_folded}
\end{figure}

As a final example, we consider a potential 4D toy realization of a 5D model with folded supersymmetry \cite{Burdman:2006tz}. In particular, we are interested in theories with a spontaneously broken $\mathbb{Z}_2$ bifold symmetry, such that the ``super-partners'' of the low energy theory have different gauge charges. For simplicity, we will just focus on a toy model that generates towers of top-like chiral multiplets, which should be dressed with electroweak gauge interactions and couplings to the Higgs to create a complete theory. Although the complete construction is detailed in \cite{Burdman:2006tz}, to guide the deconstruction we sketch the specific boundary conditions required for folded supersymmetry in Appendix \ref{app:c}.  In terms of quiver theories, the $\mathbb{Z}_2$ bifold symmetry is realized as a transformation relating two quivers, as shown in Figure~\ref{fig:quiver_folded}.

The bifold symmetry exchanges the primed gauged fields and matter content with the unprimed ones, except for the nodes $SU(k)_0$ and $SU(k)_N$ which do not transform under the  $\mathbb{Z}_2$. In order to have two separate gauge groups $SU(k)$ and $SU(k)'$ in the low energy theory, we need to break the quiver. In particular, the choice $\langle Q'_0 \rangle = \langle Q'_{\Nm{1}} \rangle = 0$ provides the right boundary conditions to yield two towers of vector modes.\footnote{This asymmetry of vevs amounts to a spontaneous breaking of the $\mathbb{Z}_2$ symmetry and could be dynamically engineered by $\mathbb{Z}_2$-breaking soft terms for the link fields, ensuring that the $\mathbb{Z}_2$ is only softly broken.}

To produce the spectrum of folded supersymmetry with light chiral fermions charged under the un-primed $SU(k)$ group, we attach an extra fundamental (anti-fundamental) at site $SU(k)_0$ ($SU(k)_N$). 

Without supersymmetry breaking terms,  the primed matter fields are all massive. To obtain a light scalar for the $SU(k)'$ gauge group, we can add a universal ``radion'' $F$-term as described in Section~\ref{subsec:lightscalar}. Such an $F$-term will split the spectrum of massive scalars to induce a light scalar mode. As detailed in the previous subsection, the choice $F_T = 2N\sin \left[ \pi/\left(4N+4\right)\right]/\pi$ results in a massless complex scalar at tree level. We have various options for inducing a top-Higgs yukawa coupling. In the 5D realization of folded supersymmetry the Higgs is a brane-localized field; in this case that would correspond to coupling Higgs multiplets in a $\mathbb{Z}_2$-symmetric fashion to two nodes, e.g., $SU(k)_1$ and $SU(k)_1'$. The yukawa coupling inherited by the zero modes in this case would be volume-suppressed, as in the 5D case, and the zero-mode scalar in the primed sector would serve as the lightest top partner. Alternately, one could imagine coupling Higgs multiplets to multiple sites, exploiting the parametric freedom of the deconstruction; we leave a detailed study to future work.


Due to the tuning of the radion term, one might worry that this deconstruction does not evidently improve upon the 5D picture, since in 5D the folded spectrum of zero modes is nominally dictated by symmetries and boundary conditions, while in the deconstruction it is dictated by the adjustment of $F$-terms.  However, even in 5D the choice of $\alpha$ required for the folded spectrum amounts to the tuning of a continuous parameter to obtain a massless scalar zero mode, entirely analogous to the tuning of F-terms in 4D. Thus the quiver presented in Figure~\ref{fig:quiver_folded} leads to a deconstruction of folded supersymmetry as expected, subject to the usual tuning of couplings required by a deconstruction. Moreover, it allows improvement over the 5D realization of folded supersymmetry, since we can realize towers of QCD and folded QCD states without demanding that all fields transform as 5D multiplets.

\section{Conclusions} \label{sec:conc}

As the LHC continues to challenge conventional supersymmetric extensions of the Standard Model, more exotic scenarios -- such as supersymmetry breaking by boundary conditions on an extra dimension -- become increasingly compelling. 
Such exotic scenarios are particularly attractive when realized in a four-dimensional framework, where the key features of extra-dimensional physics can be reproduced in a minimal setting.  In this work we have developed a successful prescription for reproducing the physics of Scherk-Schwarz supersymmetry breaking in a purely four-dimensional framework. In contrast to previous attempts, our deconstruction accommodates arbitrary Scherk-Schwarz twists and relies only on soft supersymmetry breaking, manifestly preserving supersymmetry in the ultraviolet. This completes the program of deconstructing the physics of five-dimensional supersymmetric theories compactified on an $S^1/\mathbb{Z}_2$ orbifold.

While the precise deconstruction required for a literal reproduction of the Scherk-Schwarz spectrum is somewhat baroque (a property common to all deconstructions), it  illustrates the salient details and enables the development of a wide range of simpler models with similar features. To this end, we have successfully reproduced the spectra of several specific 5D models, including the bulk Higgs construction of \cite{Barbieri:2001yz} and folded supersymmetry \cite{Burdman:2006tz}. The case of folded supersymmetry is particularly interesting, as it opens the door to simpler four-dimensional constructions in which the lightest top partners are neutral under QCD.

Our results suggest a number of interesting future directions. Clearly, there are a variety of 5D models using SSSB that can now be fully reproduced in four dimensions.  While we have focused on the precise quivers and couplings required to reproduce the exact Scherk-Schwarz tree-level spectrum, simpler quivers with more generic couplings retain many of the same qualitative features, and it would be useful to explore the simplest such models in four dimensions. Although we have focused on reproducing the physics of a flat fifth dimension, it is also possible to deconstruct a warped extra dimension \cite{Randall:2002qr}. Given that complete breaking of supersymmetry by Scherk-Schwarz boundary conditions is not possible in a warped extra dimension, it would be quite compelling to construct four-dimensional models exhibiting both the qualitative effects of warping and complete Scherk-Schwarz supersymmetry breaking. Finally, we have restricted our attention to reproducing the tree-level spectrum of Scherk-Schwarz supersymmetry breaking and leave a detailed study of radiative corrections to future work. Radiative corrections in the deconstruction will not induce quadratic sensitivity to the cutoff since 4D supersymmetry is softly broken, but in general the radiative corrections to a finite quiver differ from the 5D case due to the truncation of the KK tower and different spectrum of heavy modes.

\acknowledgments

We are grateful to Nima Arkani-Hamed, Zackaria Chacko, Csaba Cs\'{a}ki, Savas Dimopoulos, Roni Harnik, Graham Kribs, and John March-Russell for useful conversations. We particularly thank David E.~Kaplan for helpful insights regarding the relation between Scherk-Schwarz supersymmetry breaking and radion mediation in five dimensions. NC is supported by the DOE under grants DOE-SC0010008, DOE-ARRA-SC0003883, and DOE-DE-SC0007897.

\appendix

\section{Non-abelian deconstruction} \label{app:a}
In this Appendix we present the details of the matching between the 5D action and the 4D deconstructed action in the large $N$ limit for a non-abelian gauge theory. The action for the circular quiver with non-abelian nodes is 
\begin{align}
S=\int d^4x \, \left( 
\int d^2 \theta \sum_i^N \frac{1}{g^2} \tr\, W^{\alpha i} W_{\alpha i} + \textrm{h.c.} +
\int d^4 \theta \sum_i^N  \tr\, Q_i^\dagger e^{V_i} Q_i e^{-V_{i+1}}
\right)
\end{align}
The additional complication relative to the abelian case is that all fields are matrices and they do not simply commute with each other. The definitions for the variables we use are listed in Table \ref{table:dictionary}. To proceed further, we expand $Q_i =v\left(1+\epsilon R\chi_i/\sqrt{2}\right)$ and $V_{i+1}= V_{i}+ \epsilon \delta V_{i}$. For non-linear link-fields, additional $\mathcal{O}(\epsilon^2)$ terms will be present but they can be ignored when supersymmetry is unbroken. The superpotential can be directly translated to a 5D continuum limit. For the K\"{a}hler potential, we expand up to second order in $\epsilon$. The $e^{-V_{i+1}}$ is a little more complicated, and we will for now assume that $e^{-V_{i+1}} = e^{-V_{i}}+ \epsilon V_{i}^{(1)}+ \epsilon^2 V_{i}^{(2)}+ \mathcal{O}(\epsilon^3)$, where we have denoted $V_{i}^{(n)}$ as the $n^{\textrm{th}}$ order expansion of $e^{-V_{i}-\epsilon\delta V_{i}}$. The K\"{a}hler potential becomes
\begin{align}
&\int d^4 \theta \;  (v^\dagger v) \sum_i^N  \tr\,\left[
\left(1+ \frac{\epsilon R}{\sqrt{2}} \chi_i^\dagger\right)e^{V_i}
\left(1+ \frac{\epsilon R}{\sqrt{2}} \chi_i\right)\left(e^{-V_{i}}+ \epsilon V_{i}^{(1)}+ \epsilon^2 V_{i}^{(2)}\right)
 \right]  + \mathcal{O}(\epsilon^3) \notag\\
=& \int d^4 \theta \;  (v^\dagger v) \
\sum_i^N \tr\, \left[
1+\frac{\epsilon R}{\sqrt{2}} \left(\chi^\dagger_i + \chi_i\right) + \epsilon\, e^{V_i} V_i^{(1)} + 
\frac{\epsilon^2 R}{\sqrt{2}} \left(\chi^\dagger_i e^{V_i}V_i^{(1)} + e^{V_i}\chi_i V_i^{(1)} \right)\right. \notag\\
&\qquad \qquad\qquad \qquad \quad
+\left. \frac{\epsilon^2 R^2}{2} \chi_i^{\dagger}e^{V_i}\chi_i e^{-V_i} + e^{V_i}V_i^{(2)}
\right]
 +\mathcal{O}(\epsilon^3)
\end{align}
A few simplifications are possible. First let us compute the terms $V_i^{(1)}$ and $V_i^{(2)}$. To do so, it is useful to employ the systematic expansion
\begin{align}
e^{-V_{i} - \epsilon \delta V_{i}}=
-\epsilon \int_0^1 dt\, e^{-(1-t)V_i}\,\delta V_{i}\,e^{-tV_i} +
\frac{\epsilon^2}{2}\,\mathcal{T}\!\!\int_0^1\!\int_0^1 dt dt'\, e^{-(1-t)V_i}\,\delta V_{i}\,e^{-(t-t')V_i}\,\delta V_{i}\,e^{-t'V_i} + \mathcal{O}(\epsilon^3)
\end{align}
which is familiar as the standard expansion for the evolution operator in quantum mechanics, where the hamiltonian $V_i$ is perturbed by $\epsilon \delta V_{i}$. Here $\mathcal{T}$ denotes (descending) time-ordering of the integrand. The first term can be evaluated in closed form as
\begin{align}
e^{V_i}V_i^{(1)}= f(L_{V_i})(\delta V_{i}) \qquad {\rm with} \qquad
f(x)=\frac{1-e^{x}}{x}
\end{align}
where $L_{V_i}$ denotes the Lie derivative, i.e. $L_{V_i} A = [V_i, A]$. Note the well known formula $e^{A}\partial e^{-A}= f(L_A)\partial A$. In the continuum limit, we have $V_i^{(1)}\rightarrow e^{V_i}\partial_{\phi_5}e^{-V_i}$. Also note that written in terms of Lie derivatives, $e^{V_{i}}V_{i}^{(1)}$ is in the lie algebra of the gauge group and has vanishing trace. 

Let us turn to $V_{i}^{(2)}$, which appears in the form
\begin{align}
\tr\, e^{V_{i}}V_{i}^{(2)}&=\frac{1}{2}\,\tr\mathcal{T}\!\!\int_0^1\!\int_0^1 dt dt'\, e^{tV_i}\,\delta V_{i}\,e^{-(t-t')V_i}\,\delta V_{i}\,e^{-t'V_i} \notag \\
&=\tr \!\int_0^1\! dt \, e^{tV_i}\,\delta V_{i}\,e^{-tV_i}
\left(\int_0^t \! dt' e^{t'V_i}\,\delta V_{i}\,e^{-t'V_i} \, \right) 
\end{align}
Denoting $g(t)=-\int_0^t \! dt' e^{t'V_i}\,\delta V_{i}\,e^{-t'V_i} $, such that $g(1)=f(L_{V_i})\delta V_i$, he integral becomes
\begin{align}
\tr\, e^{V_{i}}V_{i}^{(2)}&=\tr \!\int_0^1\! dt \, g'(t) g(t) =\frac{1}{2}\, \tr\, g^2(1)= \frac{1}{2}\, \tr \left[f(L_{V_i})\delta V_i\right]^2
\end{align}
Altogether, the K\"{a}hler potential becomes
\begin{eqnarray}
\int d^4 \theta \;  (v^\dagger v) \
\sum_i^N \tr\, \left\{
\frac{\epsilon^2}{2} 
\left[f(L_{V_i})(\delta V_i) + \frac{R}{\sqrt{2}}\left(\chi^{\dagger}_i + e^{V_i}\chi_i^{\dagger}e^{-V_i}\right)\right]^2 \right. \notag \\
\left. + \frac{\epsilon R}{\sqrt{2}} \left(\chi^\dagger_i + \chi_i\right) -
\frac{\epsilon^2 R^2}{4} \left(\chi^{\dagger 2}_i + \chi_i^2\right) 
\right\} +\mathcal{O}(\epsilon^3)
\end{eqnarray}
Similar to the $U(1)$ story, for a constant $v$, the $d^4\theta$ integration of pure chiral fields vanishes and the precise 5D action is reproduced in the continuum limit.

\section{Mass Spectra} 
\label{app:b}
In this Appendix we derive the tree level mass spectrum for quiver theories with radion mediation added. As we have seen in Section~\ref{sec:decon}, deconstructed theories reproduce the action of an extra dimension on a lattice. We then expect the equation of motion and the mass spectrum to be a latticized version of the continuum limit. In this section, we will diagonalize the mass matrix and see how various boundary conditions and Scherk-Schwarz twists are reproduced in this limit. Since the radion expectation value only alters the spectrum of gauge fermions and matter scalars, we will focus on these towers of states; the mass spectra of gauge bosons and matter fermions can be extracted from the supersymmetric limit. 

Before going through the derivation, let us define some notation to make the continuum analogue transparent. First let $(\bm{\Psi})_i=(\Psi_1, \Psi_2)_i$ be a vector in an infinite dimensional space, with $i$ ranging from $-\infty$ to $\infty$. For a vector to represent a mass eigenstates for a quiver, i.e. $\bm{\lambda}_i=\bm{\Psi}_i$, we only demand $\bm{\Psi}_i$ to be a mass eigenstate for the index $i$ such that the corresponding field is present. Obviously there are many redundancies for choosing a $\bm{\Psi}_i$ to represent a physical state. We will fix this choice later in a way that will make the boundary conditions transparent. We define the operators $\Delta$ and $f$ on these vectors in the following way

\begin{align}
(\Delta \Psi)_i = \frac{R}{a}\binom{\Psi_{1, i+1}-\Psi_{1,i}}{\Psi_{2,i}-\Psi_{2,i-1}} 
\qquad
(f \Psi)_i = \frac{1}{2}\binom{-iF_T \Psi_{2,i}}{iF_T \Psi_{1,i}}
\end{align}
Considering $\Delta$ and $f$ as infinite dimensional matrices, their conjugates can be computed

\begin{align}
(\Delta^\dagger \Psi)_i = -\frac{R}{a}\binom{\Psi_{1, i}-\Psi_{1,i-1}}{\Psi_{2,i+1}-\Psi_{2,i}} 
\qquad
(f^\dagger \Psi)_i = \frac{1}{2}\binom{-iF_T \Psi_{2,i}}{iF_T \Psi_{1,i}}
\end{align}

To see why these operators are useful, consider for a moment an infinite linear quiver. The masses come from Yukawa couplings of the $\chi_i$ after expanding around $\langle Q_i\rangle=v$,
\begin{align}
\mathcal{L}_{m_{\textrm{gaugino}}} \sim \sum_i \lambda_i^a \left[\psi^{a}_{\chi_{i}} - \psi^{a}_{\chi_{i-1}}\right] + \textrm{c.c.} \, ,
\end{align}
where we have decomposed $\psi_{\chi_i}=T^a\psi^a_{\chi_i}$ and ignored the $U(1)$ components. In the continuum limit, terms of the form $\sum_i\lambda_i (\psi^{a}_{\chi_{i+1}} - \psi^{a}_{\chi_{i}})$  become $\lambda \partial_\phi \psi_\chi$. However, the discrete derivatives for $\lambda_i$ and $\psi^{a}_{\chi_{i}}$ differ, as is made apparent by rewriting the gaugino mass terms in the form
\begin{align}
\sum_i  \lambda_i^a \left(\psi^{a}_{\chi_{i}} - \psi^{a}_{\chi_{i-1}}\right) = 
-\sum_i  \left(\lambda_{i+1}^a - \lambda_{i}^a \right) \psi^{a}_{\chi_i}
\end{align}
In order to produce the correct mass terms, the corresponding discrete derivatives for $\psi_{\chi_i}$ and $\lambda_i$ have to be $\Delta \psi_{\chi_i} \sim \psi_{\chi_{i}} - \psi_{\chi_{i-1}}$ and $\Delta\lambda_i \sim \lambda_{i+1} - \lambda_{i}$ respectively. We can then define the discrete derivative for $\bm{\lambda}_j=(\lambda, -i\psi_{\chi})_j$ by
\begin{align}
(\Delta \bm{\lambda})_j= \frac{R}{a}\binom{\lambda_{j+1} - \lambda_{j}}{-i(\psi_{\chi_{j}} - \psi_{\chi_{j-1}})}
\end{align}
The mass term can then be written in terms of the discrete derivative as
\begin{equation}
\mathcal{L} \ni 
i\bm\lambda_i^{a\dagger} \bar\sigma^\mu \partial_\mu \bm\lambda_i^a - 
\frac{1}{2}\bm\lambda_i^a M_{ij} \bm\lambda_j^a 
=
i\bm\lambda_i^{a\dagger} \bar\sigma^\mu \partial_\mu \bm\lambda_i^a + 
\frac{1}{2R}\bm\lambda_i^a \epsilon_{ik}\left(\delta_{kj}\Delta +i f_{kj}\right) \bm\lambda_j^a 
\qquad
f_{ij}=
\frac{1}{2}
\left(
\begin{matrix}
0 & -iF_T \\
iF_T & 0 
\end{matrix}
\right)
\label{eq:mass_matrix}
\end{equation}
To obtain the physical masses, we diagonalize via Takagi factorization and rewrite $M=U^TDU$ where $D$ is diagonal and $U$ is unitary. The eigenstates and eigenvalues can be obtained by finding the eigenstates and the square root of the eigenvalues of $M^\dagger M=(\Delta + if)^\dagger(\Delta + if)$. Since the quiver is translationally invariant, we can solve for the eigenstates $\bm\eta_p$ by
\begin{align}
(\bm{\eta}_{p})_j= e^{ipa j}\bm{\eta}_{p0}
\qquad
\left|
\left(
\begin{matrix}
e^{ipa}-1 & \frac{aF_T}{2R} \\
-\frac{aF_T}{2R} & 1-e^{-ipa}
\end{matrix}
\right)
\right|^2 \bm{\eta}_{p0} = m_p^2 \, \bm{\eta}_{p0}
\label{eq:app_eigenstates}
\end{align}
There are two sets of solutions for $\bm{\eta}_0$ and $m$, corresponding to left- and right- moving modes:
\begin{align}
\bm{\eta}_0^{\pm} = \binom{e^{-\frac{ipa}{2}}}{\pm i}  \qquad
m_p^{\pm} = \left(\frac{2}{a}\right) \left| \sin \left( \frac{pa}{2} \right)\pm \frac{a F_T}{4R} \right| 
\end{align}
For the infinite quiver, the mass eigenstates are then labeled by the momentum $p$ and helicity $\pm$, with no restrictions on the momentum $p$. An infinite quiver corresponds to the limit $a\to 0$ and a continuum tower is recovered. Matter can be added to the quiver, and the mass terms are exactly analogous:
\begin{align}
\mathcal{L}\ni \frac{1}{R^2}\left|(\Delta + if)\bm{\varphi}\right|^2
\end{align}
where $(\bm{\varphi})_j = (\varphi_i,\varphi_i^{c\dagger})$. The mass eigenstates can be computed by directly diagonalizing $M^\dagger M$, and the solution is the same as the gauginos.

\subsection{Circular Quiver}
Using the solution in Equation \ref{eq:app_eigenstates}, the eigenstates and mass spectrum for finite quivers can be derived. For the case of a circular quiver, the mass terms in Equation \ref{eq:mass_matrix} for the bulk of the quiver are still the same. However, since $\bm{\lambda}_i$ are not physical for $i\notin (0,...,N-1)$, the $\Delta$ operator needs to be modified. The same modification also applies for additional matter as well. We denote $\qDelta$, $\qf$ and $\qM = \qDelta + i\qf$ as the modified operators. For the circular quiver, $\qf = f$, but $\qDelta$ is modified at the boundary

\begin{align}
(\qDelta \bm{\Psi})_0= \frac{R}{a}\binom{\Psi_{1, 1}-\Psi_{1,0}}{\Psi_{2,0}-\Psi_{2,N-1}} 
\qquad 
(\qDelta \bm{\Psi})_{N-1}= \frac{R}{a}\binom{\Psi_{1, 0}-\Psi_{1,N-1}}{\Psi_{2,N-1}-\Psi_{2,N-2}}
\end{align}
We can fix an ansatz for the eigenstates $\bm\Psi_j = (\bm{\eta}_p)_j$ from Equation \ref{eq:app_eigenstates}, such that the mass eigenstate requirement becomes

\begin{align}
(\qM^\dagger \qM \bm\Psi)_j=
\big[(\qDelta + i\qf)^\dagger (\qDelta+i\qf) \bm{\Psi}\big]_j = m_p^2  \bm{\Psi}_j = 
\big[(\Delta + if)^\dagger (\Delta+if) \bm{\Psi}\big]_j= (M^\dagger M \bm\Psi)_j
\end{align}
By using the momentum and helicity eigenstates as ans\"{a}tze, finding the mass eigenstates is reduced to matching the operators $\qM^\dagger\qM$ and $M^\dagger M$ at the boundary. For the circular quiver, the matching can be easily done by demanding a periodic boundary condition $(\bm{\eta}_p)_{j+N}=(\bm{\eta}_p)_{j}$, and we arrive at the spectrum,
\begin{align}
\bm{\lambda}_{nj}^{\pm}= \bm{\eta}^\pm_{p_n}(j) \qquad
p_na=\frac{2\pi n}{N}&= \frac{an}{R} 
\qquad
m_n^{\pm} = \left(\frac{2}{a}\right) \left| \sin \left( \frac{na}{2R} \right)\pm \frac{a F_T}{4R}\right|
\qquad
n \in (0, 1, ..., N-1) 
\end{align}

\subsection{Orbifold Quiver}

The same procedure can be applied to the orbifold case. Let us first consider a semi-infinite quiver where all fields with indices $i<0$ are unphysical. Unlike the circular quiver, the translational symmetry $i\rightarrow i+1$ is broken, and we have to assume a more general ansatz 
\begin{align}
\bm{\Psi}_j =
\big(c^+ \bm{\eta}^+_{p}+c^-\bm{\eta}^-_{-p}\big)_j
\end{align}
where the $c^\pm$ are some fixed constants to be determined later when fixing the boundary conditions. The ansatz satisfies $M^\dagger M \bm\Psi_j = m_p^2 \bm\Psi_j$ since the states $\bm\eta^+_p$ and $\bm\eta^-_{-p}$ are degenerate. The discrete derivative for this quiver, $\qDelta$, has to be modified at the boundary,

\begin{align}
(\qDelta \bm\Psi)_{1,j}= \frac{R}{a}
\begin{cases}
\Psi_{1,j+1} - \Psi_{1,j} &\quad j \ge 0 \\
0 &\quad \textrm{else}
\end{cases}
\qquad
(\qDelta \bm\Psi)_{2,j}= \frac{R}{a}
\begin{cases}
\Psi_{2,j} - \Psi_{2,j-1} &\quad j \ge 1 \\
\Psi_{2,0} &\quad j=0 \\
0 &\quad \textrm{else}
\end{cases}
\label{eq:orbifold_difference}
\end{align}
The modification for $\qDelta^\dagger$ is entirely analogous,
\begin{align}
(\qDelta^\dagger \bm\Psi)_{1,j}= -\frac{R}{a}
\begin{cases}
\Psi_{1,j} - \Psi_{1,j-1} &\quad j\ge 1 \\
\Psi_{1,0} &\quad j=0 \\
0 &\quad \textrm{else}
\end{cases}
\qquad
(\qDelta^\dagger \bm\Psi)_{2,j}= -\frac{R}{a}
\begin{cases}
\Psi_{2,j+1} - \Psi_{2,j} &\quad j\ge 0\\
0 &\quad \textrm{else}
\end{cases}
\end{align}
Finally, similar modifications must be made for the operator $\qf = \qf^\dagger$ as well,
\begin{align}
(\qf \bm\Psi)_{j} =  
\begin{cases}
(f\bm\Psi)_{j} &\quad j\ge 0 \\
0 &\quad \textrm{else}
\end{cases}
\quad\quad 
f\bm\Psi_{j} = \frac{iF_T}{2}\binom{-\Psi_{2,j}}{\Psi_{1,j}}
\end{align}
To calculate the spectrum, we have to match the operators $\qM^\dagger \qM$ and $M^\dagger M$ at the boundary,
\begin{align}
\big[(\qDelta + i\qf)^\dagger (\qDelta+i\qf) \bm{\Psi}\big]_0
=\big[(\qDelta^\dagger\qDelta + i\qDelta^\dagger\qf - i\qf^\dagger\qDelta + \qf^\dagger \qf)\bm{\Psi}\big]_0
\end{align}
Let us compute the above quantity term by term, first for the upper component, only $\qDelta^\dagger \qDelta$ and $\qDelta^\dagger \qf$ are modified
\begin{align}
(\qDelta^\dagger\qDelta \bm\Psi)_{1,0}&=  -\frac{R}{a}(\qDelta \bm\Psi)_{1,0} = -\frac{R^2}{a^2}(\Psi_{1,1}-\Psi_{1,0}) \\
(\qDelta^\dagger \qf \bm\Psi)_{1,0}&= -\frac{R}{a}(\qf \bm\Psi)_{1,0} = \frac{iF_TR}{2a}\Psi_{2,0}
\end{align}
For the lower component, only $\qDelta^\dagger \qDelta$ is modified
\begin{align}
(\qDelta^\dagger\qDelta \bm\Psi)_{2,0}&=  -\frac{R}{a}\left[(\qDelta \bm\Psi)_{2,1}-(\qDelta \bm\Psi)_{2,0}\right] = -\frac{R^2}{a^2}(\Psi_{2,1}-2\Psi_{2,0})
\end{align}
Matching the boundary conditions $(\qM^\dagger \qM \bm\Psi)_0 = (M^\dagger M \bm\Psi)_0$, we arrive at
\begin{align}
-\frac{R^2}{a^2}(\Psi_{1,1}-\Psi_{1,0}) - \frac{F_TR}{2a}\Psi_{2,0} &=
-\frac{R^2}{a^2}(\Psi_{1,1}-2\Psi_{1,0}+\Psi_{1,-1}) - \frac{F_TR}{2a}(\Psi_{2,0}-\Psi_{2,-1})
\\
-\frac{R^2}{a^2}(\Psi_{2,1}-2\Psi_{2,0}) &= -\frac{R^2}{a^2}(\Psi_{2,1}-2\Psi_{2,0}+\Psi_{2,-1})
\end{align}
The set of linear equations can be solved, 
\begin{align}
\Psi_{1,-1} - \Psi_{1,0} = 0 \quad \textrm{and} \quad \Psi_{2,-1}=0
\end{align}
We see that for the semi-infinite quiver, the mass eigenstates have Neumann boundary condition for the field $\Psi_{1,j}$ and Dirichlet boundary condition for the field $\Psi_{2,j}$ at $j=0$. Since the labeling $j$ is arbitrary, the conclusion is valid for any boundary in the quiver. The following general statement is then true:

\begin{quote}
Align the indices such that $+$ indicates the direction toward the bulk of the quiver. Let $b$ be a boundary point such that fields $\bm\Psi_b$ are physical but fields with indices less than $b$ are unphysical. If $\bm\Psi = (\Psi_+, \Psi_-)$ such that $\Delta \Psi_{+b} \sim \Psi_{b+1}-\Psi_{b}$ and $\Delta \Psi_{-b} \sim \Psi_{b}-\Psi_{b-1}$, then $\Psi_+$ and $\Psi_-$ have Neumann and Dirichlet boundary conditions at the boundary respectively, or explicitly $\Delta \Psi_{+, b-1}=0=\Psi_{-, b-1}$
\end{quote}
 
Let us apply our derivations to various orbifold quivers. Consider a finite linear quiver with gauge nodes with indices $i\in (0, 1, ..., N-1)$ and link fields with indices $i\in (0, 1, ..., N-2)$. The boundary conditions at $N-1$ can be computed by re-indexing the quiver from the other end. The mass terms are symmetric under reflection, and thus we have
\begin{align}
\Psi_{1,0} - \Psi_{1,-1}= \Psi_{N} - \Psi_{N-1} = 0
\qquad
\Psi_{2,-1} = \Psi_{2,N-1} = 0
\end{align}
The mass eigenstates that satisfy the above boundary conditions are

\begin{align}
\bm{\Psi}_{nj}^{\pm} = \frac{1}{\sqrt{N}}
\binom{\cos\left[p_na(j+\sfrac{1}{2})\right]}{\mp\sin[p_na(j+1)]}
\qquad p_na=\frac{\pi n}{N}&= \frac{an}{R} 
\qquad
n \in (0, 1, ..., N-1) 
\notag\\
m_{n}^{\pm} = \left(\frac{2}{a}\right) \left| \sin \left( \frac{p_na}{2} \right)\pm \frac{a F_T}{2R} \right| &= 
\left(\frac{2}{a}\right) \left| \sin \left( \frac{n a}{2R} \right)\pm \frac{aF_T}{2R} \right| 
\label{eq:app_eigenstates_orbifold}
\end{align}
Note that $p_n$ now only needs to be an integer multiple of $\pi/N$ instead of $2\pi/N$ since the periodic boundary condition is replaced by Neumann/Dirichlet boundary conditions. Equation \ref{eq:app_eigenstates_orbifold} works for the gauginos and the orbifold matter with $\varphi_{N-1}^c$ removed.

Consider another quiver configuration, where the gauge and link fields are as before, but no matter fields are removed so that the matter spectrum is vector-like. Such a quiver is relevant to the deconstruction of folded supersymmetry. All the chiral fields will then pair up and become massive. We can apply the same logic as above and obtain boundary conditions:
\begin{align}
\Psi_{1,0} - \Psi_{1,-1}= \Psi_{N} = 0
\qquad
\Psi_{2,-1} = \Psi_{2,N}-\Psi_{2,N-1} = 0
\end{align}
The boundary conditions at $i=0$ remain the same but are reversed at $i=N-1$. The eigenstates and masses are
\begin{align}
\bm{\Psi}_{nj}^{\pm}= \frac{1}{\sqrt{N}}
\binom{\cos\left[p_na(j+\sfrac{1}{2})\right]}{\mp\sin\left[p_na(j+1)\right]}
\qquad p_na=\frac{\pi (2n+1)}{2N+1}&= \frac{a\left(n +\sfrac{1}{2}\right)}{R  \left(1+\sfrac{1}{N}\right)}
\qquad
n \in (0, 1, ..., N-1) 
\notag\\
m_{n}^{\pm} = \left(\frac{2}{a}\right) \left| \sin \left( \frac{p_na}{2} \right)\pm \frac{a F_T}{2R} \right| &= 
\left(\frac{2}{a}\right) \left| \sin \left[ \frac{a\left(n +\sfrac{1}{2}\right)}{2R\left(1+\sfrac{1}{N}\right)} \right]\pm \frac{aF_T}{2R} \right| 
\label{eq:app_eigenstates_matter_orbifold}
\end{align}
The continuum limit in this case corresponds to mixed Dirichlet-Neumann and Neumann-Dirichlet boundary conditions. 

\section{5D Orbifold and Scherk-Schwarz Twist} \label{app:c}

In this Appendix we review the parity assignments of an orbifold extra dimension with SSSB with an eye towards understanding the boundary conditions required by folded supersymmetry. To begin, one typically starts with the covering space of the extra dimension, taken to be $\mathbb{R}$. Orbifolding corresponds to identifying $x_5 \longleftrightarrow -x_5$. Under such an identification, different fields may be assigned different transformation properties, with
\begin{align}
\Phi(x_5)=\mathcal{Z} \Phi(-x_5)
\label{eq:app_orbifold}
\end{align}
where $\mathcal{Z}$ is a symmetry in the un-orbifolded theory. Consistency requires that $\mathcal{Z}^2$ is the identity. The real line can be further compactified by another set of identifications, $x_5 \longleftrightarrow x_5 + 2\pi R$. Again, fields can transform non-trivially under such an identification, with
\begin{align}
\Phi(x_5) = \mathcal{T} \Phi(x_5+2\pi R)
\label{eq:app_scherk_schwarz}
\end{align}
Again, $\mathcal{T}$ must be a symmetry of the original theory. If $\mathcal{T}$ is the identity, then we simply have a theory on an orbifold $S^1/\mathbb{Z}_2$. But in general, $\mathcal{T}$ can be a non-trivial twist, as it only needs to satisfy the self-consistency equation obtained by combining identifications in the following way:
\begin{align}
\Phi(x_5) = \mathcal{Z} \Phi(-x_5) = 
\mathcal{Z} \mathcal{T}\Phi(-x_5 + 2\pi R) =
\mathcal{Z} \mathcal{T} \mathcal{Z} \Phi(x_5 - 2\pi R)= 
\mathcal{Z} \mathcal{T} \mathcal{Z} \mathcal{T} \Phi(x_5)
\end{align}
In order for the orbifolding and Scherk-Schwarz twist to be compatible, we require $\mathcal{ZTZT}=1$. Alternately, we can define $\mathcal{Z}'\equiv \mathcal{ZT}$, such that $\mathcal{Z}^{\prime 2}=1$. Then $\mathcal{Z}$ and $\mathcal{Z}'$ are mirror flips with respect to the points $x_5=0$ and $x_5=\pi R$, respectively. For the purpose of deconstruction, it is useful to instead consider the restricted domain $x_5 \in [0, \pi R]$, for which imposing an even or odd parity under $\mathcal{Z},\mathcal{Z}'$ is equivalent to imposing Neumann or Dirichlet boundary conditions. This is particularly helpful to understand the boundary condition assignments in folded supersymmetry.

To understand the case of folded supersymmetry, let us focus on the bulk matter content $\Phi=(\phi, \phi^{c\dagger}, \psi, \psi^{c\dagger})$ and $\Phi'=(\phi', \phi^{c\dagger\prime}, \psi', \psi^{c\dagger\prime})$. $\Phi$ is a visible sector field and $\Phi'$ is a folded sector field, and the fields in parentheses denote the corresponding scalar and fermionic components. The appropriate set of charge assignments for a folded spectrum is
\begin{table}[h]
\begin{center}
\begin{tabular}{c|cccc|ccccc}
&
$\phi$ & $\phi^{c\dagger}$ & $\psi$ & $\psi^{c\dagger}$ &
$\phi'$ & $\phi^{c\dagger\prime}$ & $\psi'$ & $\psi^{c\dagger\prime}$ 
\\
\hline
$\mathcal{Z}$ &
$1$ & $-1$ & $1$ & $-1$ &
$1$ & $-1$ & $1$ & $-1$
\\
$\mathcal{Z}'$ &
$-1$ & $1$ & $1$ & $-1$ &
$1$ & $-1$ & $-1$ & $1$
\end{tabular}
\end{center}
\caption{Parity assignment for folded supersymmetry. \label{table:folded_charge}}
\end{table}

We see that each $\mathcal{Z}$ and $\mathcal{Z}'$ preserve a separate four-dimensional $\mathcal{N}=1$ supersymmetry, but taken together, no supersymmetry remains. In general, however, there is no reason why the eigenstates of $\mathcal{Z}$ and $\mathcal{Z}'$ have to be identical. There is a continuous choice of $\alpha$ that dictates which linear combination of $\Phi,\Phi^{c\dagger}$ are eigenstates of $\mathcal{Z}'$. The parity assignments in Table \ref{table:folded_charge} correspond to the special choice of $\alpha = \pi$. One may wonder what happens if supersymmetry breaking effects are removed. Table \ref{table:folded_charge_unbroken} shows the same orbifold and Scherk-Schwarz parity assignments with $\alpha = 0$. While supersymmetry is not broken, the bifold symmetry $\Psi' \longleftrightarrow \Psi$ is still broken by the Scherk-Schwarz Mechanism.

\begin{table}[h]
\begin{center}
\begin{tabular}{c|cccc|ccccc}
&
$\phi$ & $\phi^{c\dagger}$ & $\psi$ & $\psi^{c\dagger}$ &
$\phi'$ & $\phi^{c\dagger\prime}$ & $\psi'$ & $\psi^{c\dagger\prime}$ 
\\
\hline
$\mathcal{Z}$ &
$1$ & $-1$ & $1$ & $-1$ &
$1$ & $-1$ & $1$ & $-1$
\\
$\mathcal{Z}'$ &
$1$ & $-1$ & $1$ & $-1$ &
$-1$ & $1$ & $-1$ & $1$
\end{tabular}
\end{center}
\caption{Parity assignment for folded-supersymmetry with supersymmetry breaking terms removed. \label{table:folded_charge_unbroken}}
\end{table}

In Table \ref{table:folded_charge_unbroken}, the un-primed matter fields have either Neumann-Neumann or Dirichlet-Dirichlet boundary conditions at $x_5 = 0, \pi R$, whereas the primed matter fields have Neumann-Dirichlet or Dirichlet-Neumann boundary conditions. These boundary conditions are the starting point for deconstructing folded supersymmetry.

\bibliography{decon}{}

\providecommand{\href}[2]{#2}\begingroup\raggedright\begin{thebibliography}{10}

\bibitem{ATLAS-CONF-2014-009}
{\bf ATLAS} Collaboration, {\it {Updated coupling measurements of the Higgs
  boson with the ATLAS detector using up to 25 fb$^{-1}$ of proton-proton
  collision data}},  Tech. Rep. ATLAS-CONF-2014-009, CERN, Geneva, Mar, 2014.

\bibitem{CMS-PAS-HIG-13-005}
{\bf CMS} Collaboration, {\it {Combination of standard model Higgs boson
  searches and measurements of the properties of the new boson with a mass near
  125 GeV}},  Tech. Rep. CMS-PAS-HIG-13-005, CERN, Geneva, 2013.

\bibitem{ATLAS-CONF-2013-065}
{\bf ATLAS} Collaboration, {\it {Searches for direct scalar top pair production
  in final states with two leptons using the stransverse mass variable and a
  multivariate analysis technique in $\sqrt{s} = 8$ TeV pp collisions using
  20.3 fb$^{-1}$ of ATLAS data}},  Tech. Rep. ATLAS-CONF-2013-065, CERN,
  Geneva, Jul, 2013.

\bibitem{ATLAS-CONF-2013-037}
{\bf ATLAS} Collaboration, {\it {Search for direct top squark pair production
  in final states with one isolated lepton, jets, and missing transverse
  momentum in $\sqrt{s}=8,$TeV $pp$ collisions using 21 fb$^{-1}$ of ATLAS
  data}},  Tech. Rep. ATLAS-CONF-2013-037, CERN, Geneva, Mar, 2013.

\bibitem{ATLAS-CONF-2013-024}
{\bf ATLAS} Collaboration, {\it {Search for direct production of the top squark
  in the all-hadronic ttbar + etmiss final state in 21 fb-1 of p-pcollisions at
  sqrt(s)=8 TeV with the ATLAS detector}},  Tech. Rep. ATLAS-CONF-2013-024,
  CERN, Geneva, Mar, 2013.

\bibitem{Chatrchyan:2013iqa}
{\bf CMS} Collaboration, S.~Chatrchyan {\em et~al.}, {\it {Search for
  supersymmetry in pp collisions at $\sqrt{s}$ = 8 TeV in events with a single
  lepton, large jet multiplicity, and multiple b jets}},
  \href{http://xxx.lanl.gov/abs/1311.4937}{{\tt 1311.4937}}.

\bibitem{Chatrchyan:2013xna}
{\bf CMS} Collaboration, S.~Chatrchyan {\em et~al.}, {\it {Search for
  top-squark pair production in the single-lepton final state in pp collisions
  at $\sqrt{s}$ = 8 TeV}},  {\em Eur.Phys.J.} {\bf C73} (2013) 2677,
  [\href{http://xxx.lanl.gov/abs/1308.1586}{{\tt 1308.1586}}].

\bibitem{Chatrchyan:2013uxa}
{\bf CMS} Collaboration, S.~Chatrchyan {\em et~al.}, {\it {Inclusive search for
  a vector-like T quark with charge $\frac{2}{3}$ in pp collisions at
  $\sqrt{s}$ = 8 TeV}},  {\em Phys.Lett.} {\bf B729} (2014) 149--171,
  [\href{http://xxx.lanl.gov/abs/1311.7667}{{\tt 1311.7667}}].

\bibitem{Aad:2014pda}
{\bf ATLAS} Collaboration, G.~Aad {\em et~al.}, {\it {Search for supersymmetry
  at $\sqrt{s}$=8 TeV in final states with jets and two same-sign leptons or
  three leptons with the ATLAS detector}},
  \href{http://xxx.lanl.gov/abs/1404.2500}{{\tt 1404.2500}}.

\bibitem{Aad:2013wta}
{\bf ATLAS} Collaboration, G.~Aad {\em et~al.}, {\it {Search for new phenomena
  in final states with large jet multiplicities and missing transverse momentum
  at sqrt(s)=8 TeV proton-proton collisions using the ATLAS experiment}},  {\em
  JHEP} {\bf 1310} (2013) 130, [\href{http://xxx.lanl.gov/abs/1308.1841}{{\tt
  1308.1841}}].

\bibitem{Chatrchyan:2013wxa}
{\bf CMS} Collaboration, S.~Chatrchyan {\em et~al.}, {\it {Search for gluino
  mediated bottom- and top-squark production in multijet final states in pp
  collisions at 8 TeV}},  {\em Phys.Lett.} {\bf B725} (2013) 243--270,
  [\href{http://xxx.lanl.gov/abs/1305.2390}{{\tt 1305.2390}}].

\bibitem{Chatrchyan:2014lfa}
{\bf CMS} Collaboration, S.~Chatrchyan {\em et~al.}, {\it {Search for new
  physics in the multijet and missing transverse momentum final state in
  proton-proton collisions at $\sqrt{s}$ = 8 TeV}},
  \href{http://xxx.lanl.gov/abs/1402.4770}{{\tt 1402.4770}}.

\bibitem{Craig:2013cxa}
N.~Craig, {\it {The State of Supersymmetry after Run I of the LHC}},
  \href{http://xxx.lanl.gov/abs/1309.0528}{{\tt 1309.0528}}.

\bibitem{LeCompte:2011cn}
T.~J. LeCompte and S.~P. Martin, {\it {Large Hadron Collider reach for
  supersymmetric models with compressed mass spectra}},  {\em Phys.Rev.} {\bf
  D84} (2011) 015004, [\href{http://xxx.lanl.gov/abs/1105.4304}{{\tt
  1105.4304}}].

\bibitem{Kribs:2012gx}
G.~D. Kribs and A.~Martin, {\it {Supersoft Supersymmetry is Super-Safe}},  {\em
  Phys.Rev.} {\bf D85} (2012) 115014,
  [\href{http://xxx.lanl.gov/abs/1203.4821}{{\tt 1203.4821}}].

\bibitem{Kachru:1998ys}
S.~Kachru and E.~Silverstein, {\it {4-D conformal theories and strings on
  orbifolds}},  {\em Phys.Rev.Lett.} {\bf 80} (1998) 4855--4858,
  [\href{http://xxx.lanl.gov/abs/hep-th/9802183}{{\tt hep-th/9802183}}].

\bibitem{Lawrence:1998ja}
A.~E. Lawrence, N.~Nekrasov, and C.~Vafa, {\it {On conformal field theories in
  four-dimensions}},  {\em Nucl.Phys.} {\bf B533} (1998) 199--209,
  [\href{http://xxx.lanl.gov/abs/hep-th/9803015}{{\tt hep-th/9803015}}].

\bibitem{Bershadsky:1998mb}
M.~Bershadsky, Z.~Kakushadze, and C.~Vafa, {\it {String expansion as large N
  expansion of gauge theories}},  {\em Nucl.Phys.} {\bf B523} (1998) 59--72,
  [\href{http://xxx.lanl.gov/abs/hep-th/9803076}{{\tt hep-th/9803076}}].

\bibitem{Kakushadze:1998tr}
Z.~Kakushadze, {\it {Gauge theories from orientifolds and large N limit}},
  {\em Nucl.Phys.} {\bf B529} (1998) 157--179,
  [\href{http://xxx.lanl.gov/abs/hep-th/9803214}{{\tt hep-th/9803214}}].

\bibitem{Bershadsky:1998cb}
M.~Bershadsky and A.~Johansen, {\it {Large N limit of orbifold field
  theories}},  {\em Nucl.Phys.} {\bf B536} (1998) 141--148,
  [\href{http://xxx.lanl.gov/abs/hep-th/9803249}{{\tt hep-th/9803249}}].

\bibitem{Schmaltz:1998bg}
M.~Schmaltz, {\it {Duality of nonsupersymmetric large N gauge theories}},  {\em
  Phys.Rev.} {\bf D59} (1999) 105018,
  [\href{http://xxx.lanl.gov/abs/hep-th/9805218}{{\tt hep-th/9805218}}].

\bibitem{Chacko:2005pe}
Z.~Chacko, H.-S. Goh, and R.~Harnik, {\it {The Twin Higgs: Natural electroweak
  breaking from mirror symmetry}},  {\em Phys.Rev.Lett.} {\bf 96} (2006)
  231802, [\href{http://xxx.lanl.gov/abs/hep-ph/0506256}{{\tt
  hep-ph/0506256}}].

\bibitem{Burdman:2006tz}
G.~Burdman, Z.~Chacko, H.-S. Goh, and R.~Harnik, {\it {Folded supersymmetry and
  the LEP paradox}},  {\em JHEP} {\bf 0702} (2007) 009,
  [\href{http://xxx.lanl.gov/abs/hep-ph/0609152}{{\tt hep-ph/0609152}}].

\bibitem{Scherk:1978ta}
J.~Scherk and J.~H. Schwarz, {\it {Spontaneous Breaking of Supersymmetry
  Through Dimensional Reduction}},  {\em Phys.Lett.} {\bf B82} (1979) 60.

\bibitem{Pomarol:1998sd}
A.~Pomarol and M.~Quiros, {\it {The Standard model from extra dimensions}},
  {\em Phys.Lett.} {\bf B438} (1998) 255--260,
  [\href{http://xxx.lanl.gov/abs/hep-ph/9806263}{{\tt hep-ph/9806263}}].

\bibitem{Antoniadis:1998sd}
I.~Antoniadis, S.~Dimopoulos, A.~Pomarol, and M.~Quiros, {\it {Soft masses in
  theories with supersymmetry breaking by TeV compactification}},  {\em
  Nucl.Phys.} {\bf B544} (1999) 503--519,
  [\href{http://xxx.lanl.gov/abs/hep-ph/9810410}{{\tt hep-ph/9810410}}].

\bibitem{Delgado:1998qr}
A.~Delgado, A.~Pomarol, and M.~Quiros, {\it {Supersymmetry and electroweak
  breaking from extra dimensions at the TeV scale}},  {\em Phys.Rev.} {\bf D60}
  (1999) 095008, [\href{http://xxx.lanl.gov/abs/hep-ph/9812489}{{\tt
  hep-ph/9812489}}].

\bibitem{Barbieri:2000vh}
R.~Barbieri, L.~J. Hall, and Y.~Nomura, {\it {A Constrained standard model from
  a compact extra dimension}},  {\em Phys.Rev.} {\bf D63} (2001) 105007,
  [\href{http://xxx.lanl.gov/abs/hep-ph/0011311}{{\tt hep-ph/0011311}}].

\bibitem{Barbieri:2001dm}
R.~Barbieri, L.~J. Hall, and Y.~Nomura, {\it {Models of Scherk-Schwarz symmetry
  breaking in 5-D: Classification and calculability}},  {\em Nucl.Phys.} {\bf
  B624} (2002) 63--80, [\href{http://xxx.lanl.gov/abs/hep-th/0107004}{{\tt
  hep-th/0107004}}].

\bibitem{Murayama:2012jh}
H.~Murayama, Y.~Nomura, S.~Shirai, and K.~Tobioka, {\it {Compact
  Supersymmetry}},  {\em Phys.Rev.} {\bf D86} (2012) 115014,
  [\href{http://xxx.lanl.gov/abs/1206.4993}{{\tt 1206.4993}}].

\bibitem{Dimopoulos:2014aua}
S.~Dimopoulos, K.~Howe, and J.~March-Russell, {\it {Maximally Natural
  Supersymmetry}},  \href{http://xxx.lanl.gov/abs/1404.7554}{{\tt 1404.7554}}.

\bibitem{Ponton:2001hq}
E.~Ponton and E.~Poppitz, {\it {Casimir energy and radius stabilization in
  five-dimensional orbifolds and six-dimensional orbifolds}},  {\em JHEP} {\bf
  0106} (2001) 019, [\href{http://xxx.lanl.gov/abs/hep-ph/0105021}{{\tt
  hep-ph/0105021}}].

\bibitem{Ade:2014xna}
{\bf BICEP2 Collaboration} Collaboration, P.~Ade {\em et~al.}, {\it {BICEP2 I:
  Detection Of B-mode Polarization at Degree Angular Scales}},
  \href{http://xxx.lanl.gov/abs/1403.3985}{{\tt 1403.3985}}.

\bibitem{ArkaniHamed:2001ca}
N.~Arkani-Hamed, A.~G. Cohen, and H.~Georgi, {\it {(De)constructing
  dimensions}},  {\em Phys.Rev.Lett.} {\bf 86} (2001) 4757--4761,
  [\href{http://xxx.lanl.gov/abs/hep-th/0104005}{{\tt hep-th/0104005}}].

\bibitem{ArkaniHamed:2001ed}
N.~Arkani-Hamed, A.~G. Cohen, and H.~Georgi, {\it {Twisted supersymmetry and
  the topology of theory space}},  {\em JHEP} {\bf 0207} (2002) 020,
  [\href{http://xxx.lanl.gov/abs/hep-th/0109082}{{\tt hep-th/0109082}}].

\bibitem{Csaki:2001em}
C.~Csaki, J.~Erlich, C.~Grojean, and G.~D. Kribs, {\it {4-D constructions of
  supersymmetric extra dimensions and gaugino mediation}},  {\em Phys.Rev.}
  {\bf D65} (2002) 015003, [\href{http://xxx.lanl.gov/abs/hep-ph/0106044}{{\tt
  hep-ph/0106044}}].

\bibitem{Csaki:2001qm}
C.~Csaki, G.~D. Kribs, and J.~Terning, {\it {4-D models of Scherk-Schwarz GUT
  breaking via deconstruction}},  {\em Phys.Rev.} {\bf D65} (2002) 015004,
  [\href{http://xxx.lanl.gov/abs/hep-ph/0107266}{{\tt hep-ph/0107266}}].

\bibitem{Cheng:2001an}
H.~Cheng, D.~Kaplan, M.~Schmaltz, and W.~Skiba, {\it {Deconstructing gaugino
  mediation}},  {\em Phys.Lett.} {\bf B515} (2001) 395--399,
  [\href{http://xxx.lanl.gov/abs/hep-ph/0106098}{{\tt hep-ph/0106098}}].

\bibitem{Kobayashi:2001fr}
T.~Kobayashi, N.~Maru, and K.~Yoshioka, {\it {4-D construction of bulk
  supersymmetry breaking}},  {\em Eur.Phys.J.} {\bf C29} (2003) 277--284,
  [\href{http://xxx.lanl.gov/abs/hep-ph/0110117}{{\tt hep-ph/0110117}}].

\bibitem{Falkowski:2002gx}
A.~Falkowski, C.~Grojean, and S.~Pokorski, {\it {Soft electroweak breaking from
  hard supersymmetry breaking}},  {\em Phys.Lett.} {\bf B535} (2002) 258--270,
  [\href{http://xxx.lanl.gov/abs/hep-ph/0203033}{{\tt hep-ph/0203033}}].

\bibitem{Iqbal:2002ep}
A.~Iqbal and V.~S. Kaplunovsky, {\it {Quantum deconstruction of a 5-D SYM and
  its moduli space}},  {\em JHEP} {\bf 0405} (2004) 013,
  [\href{http://xxx.lanl.gov/abs/hep-th/0212098}{{\tt hep-th/0212098}}].

\bibitem{Cohen:2003xe}
A.~G. Cohen, D.~B. Kaplan, E.~Katz, and M.~Unsal, {\it {Supersymmetry on a
  Euclidean space-time lattice. 1. A Target theory with four supercharges}},
  {\em JHEP} {\bf 0308} (2003) 024,
  [\href{http://xxx.lanl.gov/abs/hep-lat/0302017}{{\tt hep-lat/0302017}}].

\bibitem{Dudas:2003iq}
E.~Dudas, A.~Falkowski, and S.~Pokorski, {\it {Deconstructed U(1) and
  supersymmetry breaking}},  {\em Phys.Lett.} {\bf B568} (2003) 281--290,
  [\href{http://xxx.lanl.gov/abs/hep-th/0303155}{{\tt hep-th/0303155}}].

\bibitem{DiNapoli:2006kc}
E.~Di~Napoli and V.~S. Kaplunovsky, {\it {Quantum deconstruction of 5D SQCD}},
  {\em JHEP} {\bf 0703} (2007) 092,
  [\href{http://xxx.lanl.gov/abs/hep-th/0611085}{{\tt hep-th/0611085}}].

\bibitem{Marti:2001iw}
D.~Marti and A.~Pomarol, {\it {Supersymmetric theories with compact extra
  dimensions in N=1 superfields}},  {\em Phys.Rev.} {\bf D64} (2001) 105025,
  [\href{http://xxx.lanl.gov/abs/hep-th/0106256}{{\tt hep-th/0106256}}].

\bibitem{Kaplan:2001cg}
D.~E. Kaplan and N.~Weiner, {\it {Radion mediated supersymmetry breaking as a
  Scherk-Schwarz theory}},  \href{http://xxx.lanl.gov/abs/hep-ph/0108001}{{\tt
  hep-ph/0108001}}.

\bibitem{Barbieri:2001yz}
R.~Barbieri, L.~J. Hall, and Y.~Nomura, {\it {Softly broken supersymmetric
  desert from orbifold compactification}},  {\em Phys.Rev.} {\bf D66} (2002)
  045025, [\href{http://xxx.lanl.gov/abs/hep-ph/0106190}{{\tt
  hep-ph/0106190}}].

\bibitem{Randall:2002qr}
L.~Randall, Y.~Shadmi, and N.~Weiner, {\it {Deconstruction and gauge theories
  in AdS(5)}},  {\em JHEP} {\bf 0301} (2003) 055,
  [\href{http://xxx.lanl.gov/abs/hep-th/0208120}{{\tt hep-th/0208120}}].

\end{thebibliography}\endgroup
\bibliographystyle{jhep}

\end{document}